\pdfoutput=1
\documentclass[12pt]{iopart}

\usepackage{iopams}
\usepackage{amstext}
\usepackage{color}
\usepackage[normalem]{ulem}

\eqnobysec 

\usepackage{graphicx}

\newtheorem{definition}{\bf Definition}[section]

\usepackage{hyperref}

\begin{document}

\title[The imprint of topology on network stability]{Deciphering the imprint of topology on nonlinear dynamical network stability}

\author{
J. Nitzbon$^{1,2}$, P. Schultz$^{1,3}$, J. Heitzig$^{1}$, J. Kurths$^{1,3,4,5}$ and F. Hellmann$^{1}$ }

\address{$^{1}$Potsdam Institute for Climate Impact Research, P.O. Box 60 12 03, 14412 Potsdam, Germany}
\address{$^{2}$Faculty of Physics, University of G\"ottingen, Friedrich-Hund-Platz 1, 37077 G\"ottingen, Germany}
\address{$^{3}$Department of Physics, Humboldt University of Berlin, Newtonstr. 15, 12489 Berlin, Germany}
\address{$^{4}$Institute for Complex Systems and Mathematical Biology, University of Aberdeen, Aberdeen AB24 3UE, United Kingdom}
\address{$^{5}$Department of Control Theory, Nizhny Novgorod State University, 606950 Nizhny Novgorod, Russia}

\ead{ \mailto{nitzbon@pik-potsdam.de} } 


\begin{abstract}
Coupled oscillator networks show a complex interrelations between topological characteristics of the network and the nonlinear stability of single nodes with respect to large but realistic perturbations. We extend previous results on these relations by incorporating sampling-based measures of the transient behaviour of the system, its survivability, as well as its asymptotic behaviour, its basin stability. By combining basin stability and survivability we uncover novel, previously unknown asymptotic states with solitary, desynchronized oscillators which are rotating with a frequency different from their natural one. They occur almost exclusively after perturbations at nodes with specific topological properties.

More generally we confirm and significantly refine the results on the distinguished role tree-shaped appendices play for nonlinear stability. We find a topological classification scheme for nodes located in such appendices, that exactly separates them according to their stability properties, thus establishing a strong link between topology and dynamics. Hence, the results can be used for the identification of vulnerable nodes in power grids or other coupled oscillator networks. From this classification we can derive general design principles for resilient power grids. We find that striving for homogeneous network topologies facilitates a better performance in terms of nonlinear dynamical network stability. While the employed second-order Kuramoto-like model is parametrized to be representative for power grids, we expect these insights to transfer to other critical infrastructure systems or complex network dynamics appearing in various other fields.
\end{abstract}




\maketitle

%

\section{Introduction}
\label{sec:intro}

Many critical infrastructure and supply systems (e.g., transportation, health care or power supply) are based on structures which can be described in terms of complex networks \cite{Albert2002,Barrat2004,Boccaletti2006,Kaluza2010,KeSun2005,Newman2003}. Such real world systems often evolved for the primary goal of fulfilling a specific function while also subject to certain constraints (e.g., financing or geography). An issue which is increasingly attracting notice in various fields (or from science to policy making) is the resilience of such critical infrastructure against perturbations in the form of external shocks, internal failures or changing environmental conditions \cite{Folke2006,Gao2016,Holling1973,Mitra2015}, i.e. often non-small perturbations. Ultimately, it is highly desirable to gain a deep understanding of how optimal functionality on the one hand and resilience on the other hand can be achieved simultaneously. A still open question in this context is how the stability and resilience of networked systems is interrelated with their topological properties.

There are many concepts for assessing the stability of states within multistable dynamical systems. In classical linear stability analysis (in terms of Lyapunov exponents) the focus is on the \textit{local} properties of a system's phase space. With the concept of Master Stability Functions this approach has been very successfully transferred to networked dynamical systems \cite{Pecora1998, Barahona2002}, in particular allowing the prediction of the synchronizability of coupled oscillator networks. The concept of Lyapunov functions\cite{Belykh2004,Chiang2010,Zwillinger1997} and \textit{basin stability} in contrast present a nonlinear stability measure which also accounts for non-small perturbations and hence emphasizes a \textit{global} perspective \cite{Menck2013} on the dynamics. These concepts only account for the \textit{asymptotic} behaviour of perturbed trajectories, however, for real-world systems the \textit{transient} behaviour can be just as relevant to ensure their proper functioning. The \textit{survivability} of a deterministic dynamical system is a suitable concept which complements both linear and asymptotic approaches \cite{Hellmann2016} by focussing on transients. A complementary approach, studying the timing aspects of transient behaviour, can be found in \cite{Kittel2016} and \cite{Houghton2010,Lai2011,Politi2010,Rosin2014}. In the context of control systems, questions of transient stability are explored in Viability theory \cite{Aubin2001, Aubin2011} and under the name of robust control.

In this study we investigate the collective dynamics of power grids which can serve as prototypical examples of critical infrastructure systems \cite{Rohden2012, Dorfler2013, Schmietendorf2014}. The employed model is  the (second order) Kuramoto model of coupled oscillators \cite{Rodrigues2016}, hence the insights are rather general and can be of relevance to other fields. In particular, we use basin stability and survivability to assess the asymptotic and transient stability of power grids against large nodal perturbations, respectively. As opposed to other approaches, these \textit{sampling-based} measures allow us to study relatively high dimensional systems, and to localize perturbations on the underlying network. These (nodal) stability measures are then related to purely topological characteristics of the respective nodes in the network. In this way we are able to identify a small number of topological classes of nodes which are characterized by similar stability properties. Remarkably, we also find a novel, previously unknown asymptotic state in the system that can only be accessed by perturbations at a particular topological class of nodes. The findings complement and extend previous results on the relationship between topological motifs and stability within power grids \cite{Menck2014,Schultz2014,Witthaut2016}, and demonstrate the power of sampling based methods for studying the properties of dynamical systems on networks.




\section{Methods}
\label{sec:methods}

\subsection{Oscillator model for nodal dynamics}
\label{sec:oscillator-model}

Many biological, chemical and technical systems of $N$ coupled oscillators can effectively be described by the Kuramoto model \cite{Acebron2005, Rodrigues2016}, which is given by the following temporal evolution law for an oscillator $i$'s phase $\phi_i$:
\begin{eqnarray}
	\dot{\phi_i}=\frac{P_i}{\alpha_i}-\frac{1}{\alpha_i}\sum_{j=1}^{N}K_{ij}\sin{\left(\phi_i-\phi_j\right)}\qquad\forall\;i=1...N,	\label{eqn:kuramoto}
\end{eqnarray}
where $P_i/\alpha$ denotes the oscillators natural frequency and $K_{ij}$ reflects the coupling strength between nodes $i$ and $j$, satisfying $K_{ij}=K_{ji}>0$ if nodes $i$ and $j$ are connected and $K_{ij}=0$ otherwise. Hence, the $K_{ij}$ define the topology of a (weighted) network of oscillators. $\alpha_i$ denotes a damping coefficient of node $i$.

In many contexts the dynamics of coupled oscillators is described more accurately when inertia is accounted for. This is obtained by the second-order Kuramoto model which additionally describes the temporal evolution of the node's frequency $\omega_i$:
\begin{eqnarray}
	\dot{\phi_i}&=\omega_i\,,	\label{eqn:phi-dot}\\
	\dot{\omega_i}&=P_i-\alpha_i\omega_i-\sum_{j=1}^{N}K_{ij}\sin{\left(\phi_i-\phi_j\right)}\,.	\label{eqn:omega-dot}
\end{eqnarray}
This model was shown to effectively describe the dynamics of synchronous machines within power grids \cite{Filatrella2008,Hill2006,Machowski2008,Nishikawa2015} where the $\phi_i$ denote the phase angles and the $\omega_i$ the frequency deviations from the grid's rated frequency. In this context the $P_i$ correspond to the net power input at node $i$ which is positive (negative) for generator (consumer) buses and the $\alpha_i$ describe the strength of the electro-mechanical damping and droop control. The coupling coefficients $K_{ij}$ correspond to the capacity of the transmission lines of the power grid.



For typical parameter values, the dynamical system given by (\ref{eqn:phi-dot}) and (\ref{eqn:omega-dot}) features a stable steady state $(\phi^\ast, \omega^\ast)=(\phi^\ast_1,...\phi^\ast_N,\omega^\ast_1,...,\omega^\ast_N)$ with constant phase angles $\phi_i^\ast-\phi_j^\ast$ and vanishing frequency deviations $\omega_i^\ast=0$ at every node $i$, given by a solution the following system of nonlinear equations:
\begin{eqnarray}
	P_i&=\sum_{j=1}^{N}K_{ij}\sin{\left(\phi_i^\ast-\phi_j^\ast\right)}\qquad\forall\;i=1...N.		\label{eqn:phi-star}
\end{eqnarray}
This state corresponds to the desirable synchronous operating mode of the power grid. However, depending on the parameter choices, there might be also undesirable non-synchronous states which correspond to attracting limit cycles within the phase space of the dynamical system.

Menck et al. showed that for a single-node, connected to an infinite power grid, the limit cycle solution can be approximated as
\begin{eqnarray}
	\omega_\text{LC} (t)\approx\frac{P}{\alpha}+\frac{\alpha K}{P}\cos{\left(\frac{P}{\alpha}\,t\right)}\,,	\label{eqn:omega-LC}
\end{eqnarray}
provided that $\left|P\right|/\alpha^2\gg1$ and $\left|P\right|^2/\alpha^2\gg K$  \cite{Menck2014}.

\subsection{Random growth model for network topologies}
\label{sec:network-model}

\begin{figure}[!ht]
	\begin{indented}
	\item[]\includegraphics[width=0.6\textwidth]{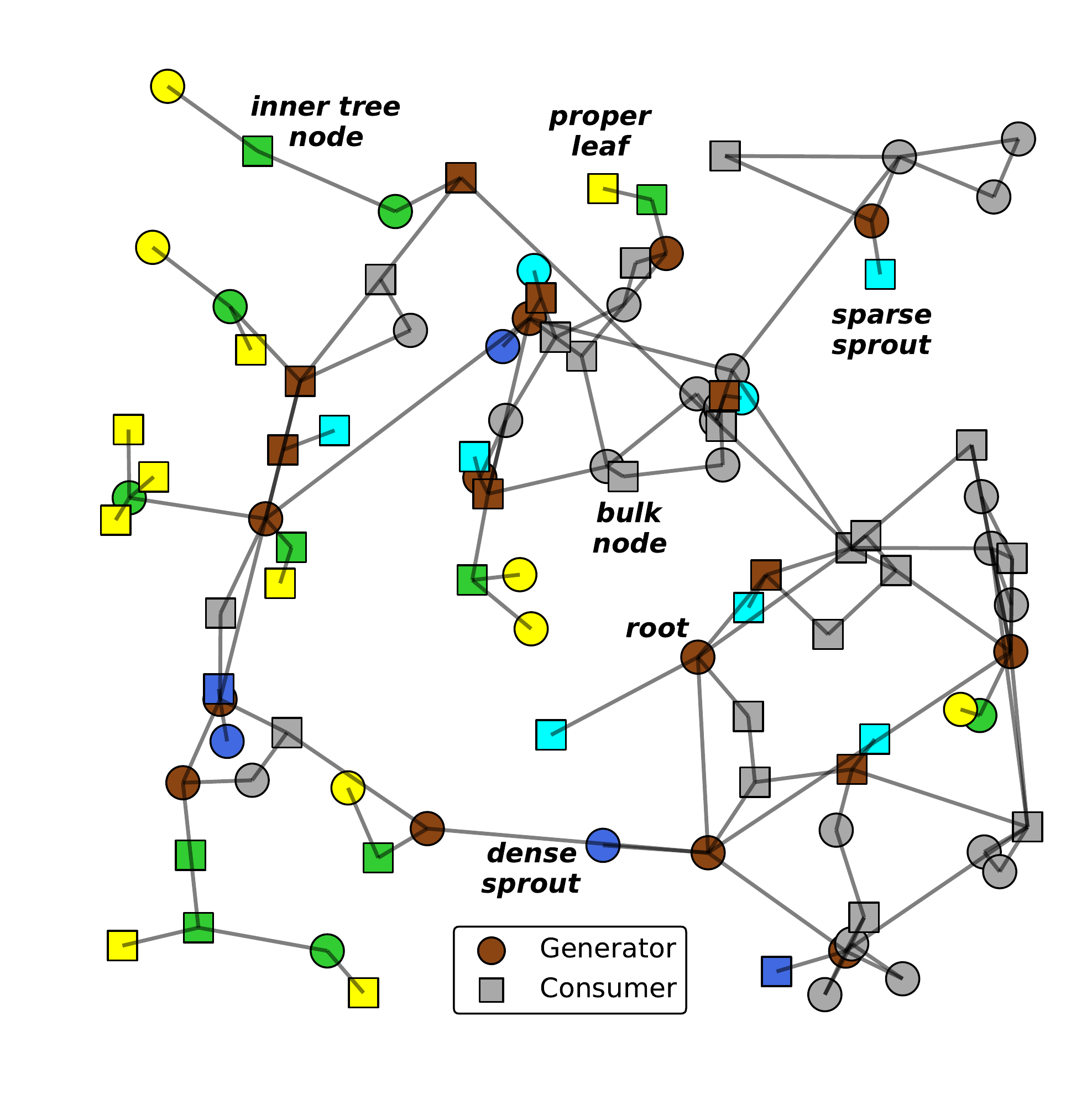}
	\end{indented}
	\caption{Spatially embedded representation of one random synthetic power grid with $N=100$ nodes of which half are net generators and half net consumers. Nodes are coloured according to their topological class, see Section \ref{sec:node-classification} for definitions.}
	\label{fig:example-network}
\end{figure}

In order to generate a representative ensemble of spatially embedded power grid topologies, we used a suitable random growth model \cite{Schultz2014}. It aims at generating synthetic network topologies that reproduce topological properties of real-world power grids and other spatially embedded infrastructure networks. The two-phase algorithm starts with a minimum spanning tree of size $N_0$ to which further nodes and (redundant) lines are added iteratively. The network growth is subject to a heuristic redundancy-versus-cost optimization, which takes not only the line lengths but also additionally-created redundancy in the form of alternative routes into account. The growth model parameters have been set to ($N_0=1$, $p=1/5$, $q=3/10$, $r=1/3$, $s=1/10$), where $p$, $q$ and $s$ are probabilities related to the creation and splitting of lines, and $r$ specifies the redundancy-versus-cost trade-off. For a detailed explanation of all parameters, we refer to \cite{Schultz2014}. For this choice of parameter values the randomly generated networks match characteristics of real-world power grids, for instance the sparsity with a mean degree of about $\bar{d}\approx2.7$. The ensemble consists of $M=50$ random networks of size $N=100$, an exemplary topology is shown in Figure \ref{fig:example-network}.

\subsection{Topological classification scheme for nodes in tree-shaped parts}
\label{sec:node-classification}

We further make use of a topological classification scheme for nodes, which particularly distinguishes nodes located in or adjacent to tree-shaped parts of the networks.
We start by giving concrete definitions for trees and tree-shaped parts and their roots:
\begin{definition}[Tree]
	A graph $G=(V,E)$ is called a \emph{tree} if it is connected and has no cycles.
\end{definition}
\begin{definition}[Tree-shaped part, root]
	Let $G=(V,E)$ be an undirected graph that is connected but not a tree. A \emph{tree-shaped part} is an induced subgraph $T^\prime=(V^\prime,E^\prime)$ of $G$ which is a tree and is maximal with the property that there is exactly one node $r\in V^\prime$ that has at least one neighbour in $G-T^\prime$. $r$ is called the \emph{root} of $T^\prime$ and has degree $d(r)\geq3$.
\end{definition}
The union of all nodes located in tree-shaped parts is subsequently denoted by $T=\bigcup_i T_i$ and called the ``forest" part of $G$, where the $T_i$ are the different tree-shaped parts within $G$. The remaining parts of $G$ are referred to as the \textit{bulk}, denoted by $B=G-T$. A finer partition of $T$ is achieved by distinguishing the root nodes $R$ from the non-root nodes $N$. The non-root nodes can be further subdivided into the leaves $L=\left\{ l\in N \,\vert\, d(l)=1 \right\}$ which have degree one, and the inner tree nodes $I=N-L$ which are located between the root and the leaves.

We will see that for stability assessment, an even finer partition is useful for whose definition we first need to introduce the following properties of nodes in tree-shaped parts:
\begin{definition}[depth, height]
	Given a tree-shaped part $T^\prime=(V^\prime,E^\prime)$ of a graph $G=(V,E)$, the \emph{depth} $\delta(x)$ of a node $x\in T^\prime$ is the length of the shortest path from $x$ to the root of $T^\prime$. The root $r\in T^\prime$ has depth $\delta(r)=0$.\\
	The \emph{height} $\eta(x)$ of a node $x\in T^\prime$ is the length of the longest outward path from $x$ to a leaf of $T^\prime$. All leaves $l\in T^\prime$ have height $\eta(l)=0$.
\end{definition}

Note that the presented definitions of height and depth of nodes might appear counter-intuitive when applied to the picture of trees growing upward from the root. However, this terminology originates from the data structure of a ``tree'' in informatics, which is typically depicted as growing downwards, and became standard in graph theory.

The smallest possible type of tree-shaped part consists of a root and some adjacent leaves. Such leaves are subsequently termed \textit{sprouts} and form the class $S=\left\{ x\in N \,\vert\, \eta(x)=0 \wedge \delta(x)=1 \right\}=\left\{ x\in L \,\vert\, \delta(x)=1 \right\}$. The leaves of larger tree-shaped parts are called \textit{proper leaves} and form the class $P=\left\{ x\in N \,\vert\, \eta(x)=0 \wedge \delta(x)>1 \right\}=\left\{ x\in L \,\vert\, \delta(x)>1 \right\}$. Note that $G=B+R+N=B+R+I+L=B+R+I+P+S$.

Finally, the group of sprouts can be separated into those which are connected to high-degree roots, called \textit{dense sprouts} $S_d=\left\{ x\in S \,\vert\, \bar{d}_\mathcal{N}>5 \right\}$, and those connected to rather low-degree roots, the \textit{sparse sprouts} $S_s=\left\{ x\in S \,\vert\, \bar{d}_\mathcal{N}<6 \right\}$ where $\bar{d}_\mathcal{N}(x)$ denotes the (average) degree of the neighbour(s) of $x$. For this last distinction, we chose the threshold of 5 so that the stability properties of the two groups are separated best.

For an efficient algorithm which provides both the partition of the nodes into the topological groups and their respective height and depth levels see \ref{sec:sm2}.

The nodes in the exemplary network in Figure \ref{fig:example-network} are coloured according to this classification and a representative node of each group is labelled accordingly. Definitions and total shares of the node categories in the ensemble of randomly generated network topologies are summarized in Table \ref{tab:overview-classification}.

\begin{table}[!ht]
	\caption{Overview of names, symbols and definitions of the hierarchically ordered topological groups of nodes in tree-shaped parts of networks. The last column shows the shares of nodes of each category in the ensemble of the $M=50$ randomly generated network topologies. More than half of the nodes belong to tree-shaped structures within the networks and about a quarter is given by leaf nodes. An exemplary network topology with the nodes coloured according to these groups is shown in Figure \ref{fig:example-network}. The simulation results shown in Figure \ref{fig:SV-vs-BS-grouped} are also coloured according to this scheme.}
	\label{tab:overview-classification}
	\begin{indented}
	\item[]\begin{tabular}{lrrr}
		\br
		Group		& Symbol	& Definition													& Share of all nodes	\\
		\br
		Bulk nodes	& $B$		& $\left\{ x\in G \,\vert\, x\notin T \right\}$					& $48.0\,\%$			\\
		Roots		& $R$		& $\left\{ x\in T \,\vert\, \exists\, b\in B: (x,b)\in E \right\}$	& $19.6\,\%$		\\
		Non-Roots	& $N$		& $\left\{ x\in T \,\vert\, x\notin R \right\}$					& $32.4\,\%$			\\
		\mr
		Inner tree nodes & $I$	& $\left\{ x\in N \,\vert\, d(x)>1 \right\}$					& $7.2\,\%$				\\
		Leaves		& $L$		& $\left\{ x\in N \,\vert\, d(x)=1 \right\}$					& $25.2\,\%$			\\
		\mr
		Proper Leaves	& $P$	& $\left\{ x\in L \,\vert\, \delta(x)>1 \right\}$				& $7.1\,\%$				\\
		Sprouts			& $S$	& $\left\{ x\in L \,\vert\, \delta(x)=1 \right\}$				& $18.1\,\%$			\\
		\mr
		Sparse Sprouts	& $S_s$	& $\left\{ x\in S \,\vert\, \bar{d}_\mathcal{N}(x)<6 \right\}$	& $12.9\,\%$			\\
		Dense Sprouts	& $S_d$	& $\left\{ x\in S \,\vert\, \bar{d}_\mathcal{N}(x)>5 \right\}$	& $5.2\,\%$				\\
		\br
	\end{tabular}
	\end{indented}
\end{table}

\subsection{Stability measures}


The subsequently introduced measures assess the stability of a dynamical system with respect to large perturbations. They reflect \textit{global} characteristics of the system's phase space which distinguishes them from the \textit{local} perspective taken in conventional linear stability analysis in which only small (infinitesimal) perturbations are regarded \cite{Pecora1998,Sun2009}. These measures are particularly suited for assessing the stability of power grids, since in this context large perturbations from the normal operating state are a common threat \cite{Dobson2013}.

The first measure to be introduced, basin stability, is an indicator for the system's likelihood to \textit{asymptotically} return to a desirable state following a large perturbation \cite{Menck2013}. The second, survivability, in turn is sensitive to whether the \textit{transient} behaviour after a larger perturbation remains within a desirable region of the system's phase space \cite{Hellmann2016}.

In complex networks, it is instructive to regard only localized perturbations originating at a single node which makes the stability measures node-wise quantities. However, it should be noted that for the stability assessment the response of the whole system with all nodes is relevant and hence the violation of the stability constraints might happen at other nodes than the perturbed one.


Both measures necessitate the specification of a probability distribution from which large (finite) perturbations are drawn. For the case of power grids modelled by (\ref{eqn:phi-dot}) and (\ref{eqn:omega-dot}) these are given by values chosen uniformly at random $(\delta\phi,\delta\omega)\in\left[-\pi,\pi\right]\times\left[-\Delta\omega,\Delta\omega \right]$ which are added 
at $t_0=0$ to the state variables of a single node $j$ while all unperturbed nodes are initialized to the desirable steady state:
\begin{eqnarray}
	\phi_i(0)&=\phi_i^\ast+\delta_{ij}\delta\phi\,,	\label{eqn:phi-pert}\\
	\omega_i(0)&=\delta_{ij}\delta\omega\,,	\label{eqn:omega-pert}
\end{eqnarray}
where $\delta_{ij}$ denotes the Kronecker delta. Recall that $\omega_i^\ast = 0$.

\begin{figure}[!ht]
	\centering
	\includegraphics[width=0.9\textwidth]{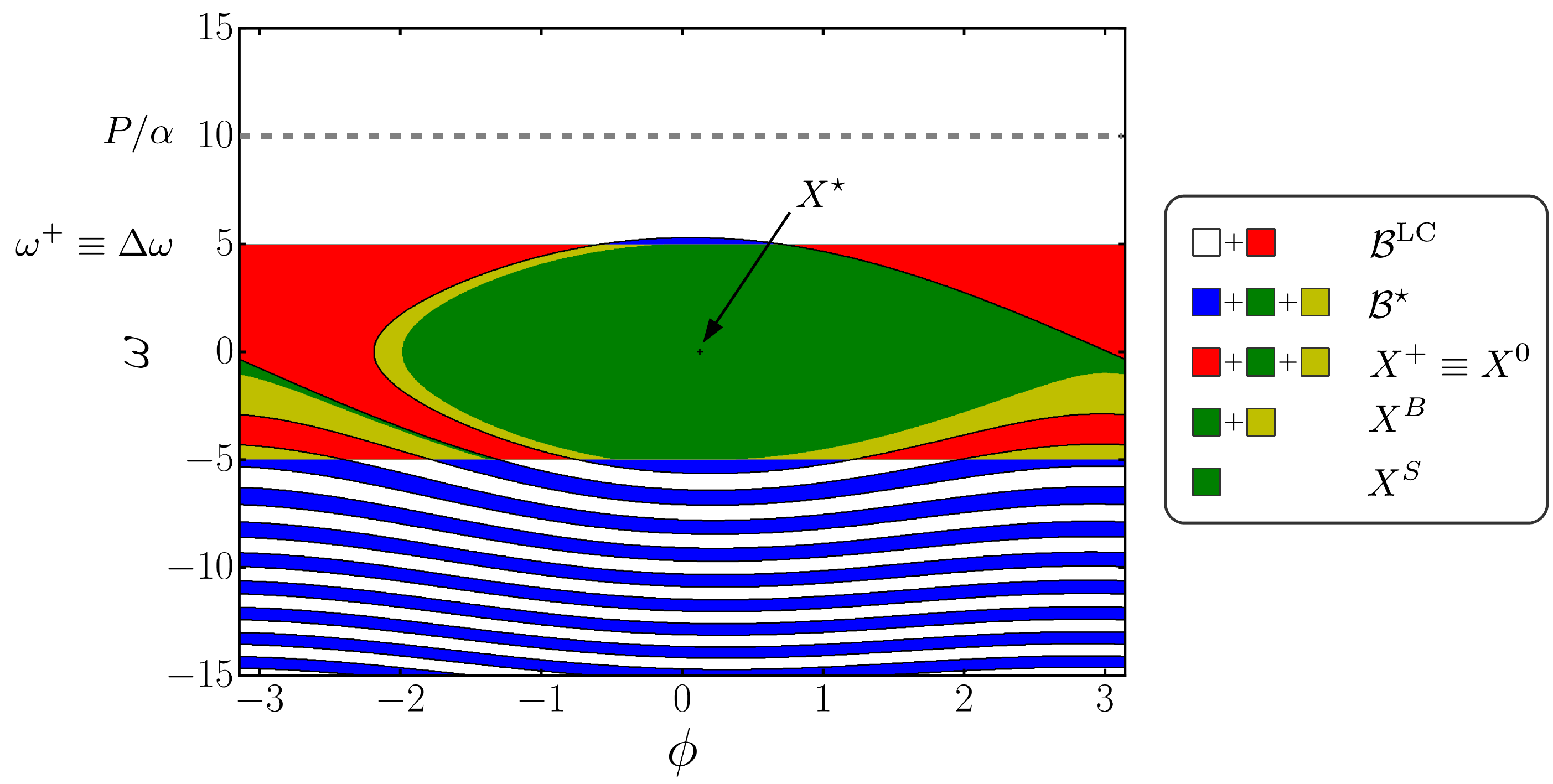}
	\caption[]{Schematic of the single-node model's phase space.
	The union of the blue, green and yellow areas is the synchronous state's ($X^\star$) basin of attraction $\mathcal{B}^\star$ while trajectories starting from the remaining parts of the phase space ($\mathcal{B}^\text{LC}$; red and white) converge to the (non-synchronous) limit cycle located around $\omega=P/\alpha$ (grey dashed). The union of the red, green and yellow areas forms the subset $X^0$ from which the random perturbations are drawn and (here) coincides with the desirable region $X^+$ relevant for the survivability measure. The green coloured region shows the (infinite-time) basin of survival $X^S$. While trajectories starting within the yellow region converge to the synchronous operating state and are thus asymptotically stable their transient leaves the desirable region $|\omega|\leq\omega^+$ and hence do not ``survive'' the perturbation.}
	\label{fig:phase-space}
\end{figure}

\subsubsection{Basin stability}
\label{sec:basin-stability}
The concept of basin stability is applicable to multistable dynamical systems with state space $X$ for which there exist attracting states distinct from the set of desirable attracting states, where the latter is denoted by $X^\star\subset X$ in the following. The \textit{basin of attraction} $\mathcal{B}^\star$ of $X^\star$ is given by all initial states from which the system asymptotically converges to the desirable attractor:
\begin{eqnarray}
	\mathcal{B}^\star=\left\{ x(0)\in X \,\left\vert\, \lim_{t\rightarrow\infty} x(t)\in X^\star \right.\right\}\,.	\label{eqn:Bstar}
\end{eqnarray}
When the perturbations are drawn from a certain subset $X^0\subseteq X$, it is instructive to define the basin of attraction restricted to this region: $X^B=\mathcal{B^\star}\cap X^0$ (cf. Figure \ref{fig:phase-space}). This is especially the case if the phase space is not compact.
In the case of perturbations drawn uniformly at random, the basin stability $\beta$ is then simply the ratio of the volume of the (restricted) basin of attraction $X^B$ to the overall region of perturbations $X^0$ \cite{Menck2013,Schultz2016}:
\begin{eqnarray}
	\beta=\frac{\text{Vol} \left( X^B \right)}{\text{Vol} \left( X^0 \right)}\,.	\label{eqn:muB}
\end{eqnarray}
In other words, this quantity corresponds to the probability for the system to return to the desirable attractor after a perturbation from $X^0$.

For the synchronous machine power grid model described by (\ref{eqn:phi-dot}) and (\ref{eqn:omega-dot}) the desirable attractor $X^\star$ is identical to the set of synchronous states $(\phi^\ast, \omega^\ast = 0)$, while there exist several non-synchronous attracting states which are undesirable. 
Since basin stability is determined for each individual node $j$, the region of perturbations is the subset
\begin{eqnarray}
	X^0_j=\{ (\phi, \omega ) \subset X \,\vert\, &-\pi \leq \phi_j \leq \pi \wedge -\Delta\omega\leq \omega_j \leq \Delta\omega  \nonumber\\
												& \wedge \,\forall\,i\neq j: ( \phi_i=\phi_i^\ast \wedge \omega_i=0 ) \} \,. \label{eqn:X0}
\end{eqnarray}
Hence the single-node basin stability corresponds to the ratio of areas in the phase space cross-section spanned by the dimensions associated to the node $j$ (cf. Figure \ref{fig:phase-space}).

As the basin of attraction and its geometry are typically not known a priori and difficult to determine, especially in high-dimensional systems \cite{Lovasz2003}, $\beta$ needs to be determined numerically via a Monte-Carlo method. For each node $L=200$ independent perturbations have been chosen uniformly at random from $X^0_j$ and the corresponding trajectories simulated for $t=100$ time units. The basin stability $\beta$ of node $j$ can be estimated from the number $s$ of trajectories which asymptotically return to $X^\star$. More details are given in \ref{sec:sm1}.

\subsubsection{Survivability}
\label{sec:survivability}
In contrast to basin stability the survivability concept\cite{Hellmann2016} presumes a desirable region $X^+\subseteq X$ which must not be left by a trajectory for a time $t$ following a large perturbation in order to call the system ``survived''. The \textit{finite-time basin of survival} $X^S_t$ is given by the fraction of those initial states of the system which give rise to evolutions that stay within $X^+$ until time $t$:
\begin{eqnarray}
	X^S_t=\left\{ \left. x(0)\in X^0 \,\right\vert\, x(t^\prime)\in X^+\; \forall \; 0\leq t^\prime \leq t \right\}\,.	\label{eqn:XSt}
\end{eqnarray}
The \textit{infinite-time basin of survival} is obtained by taking the limit $X^S=\lim_{t\rightarrow\infty} X^S_t$. The survivability $\sigma$ of a system with respect to uniformly drawn random perturbations from the region $X^0$ is then analogously given as the ratio  (cf. Figure \ref{fig:phase-space})
\begin{eqnarray}
	\sigma=\frac{\text{Vol} \left( X^S \right)}{\text{Vol}\left( X^0 \right)}\,.		\label{eqn:muS}
\end{eqnarray}
Hence it can be regarded as the probability for the system to remain within a desirable region $X^+$ after being hit by a perturbation from $X^0$.

For the operation of power grids it is necessary to keep the frequency deviations of all generators and consumers below a certain level \cite{ENTSO-E2013}. Hence for the synchronous machine model the desirable region of the state space is given as 
\begin{eqnarray}
	X^+=\left\{ (\phi_1,...,\phi_N,\omega_1,...,\omega_N)\subseteq X \,\vert\; \forall\,i:\,|\omega_i|\leq \omega^+ \right\}\,,	\label{eqn:X+}
\end{eqnarray}
with $\omega^+>0$ being the maximally tolerable frequency deviation. Again for each node $j$ the perturbations are chosen at random from $X^0_j$ (cf. (\ref{eqn:X0})) and the basin of survival is determined node-wise as $X^S_j\subseteq X^0_j$ (Figure \ref{fig:phase-space}). Note that it is not of relevance at which node the frequency constraint $|\omega|\leq\omega^+$ is violated for the perturbation to count as not survived. Subsequently, the desirable region boundaries are chosen identical to the maximal perturbation level of the frequency deviations, $\omega^+\equiv\Delta\omega$.

In order to estimate the single-node survivability $\sigma_j$ for each node, $L=200$ trajectories with independent perturbations to the synchronous operating state, drawn uniformly from $X^0_j$ (cf. \ref{eqn:X0}), have been simulated. The fraction of trajectories which did not leave the desirable region $X^+$ has been used as a statistical estimator for $\sigma_j$ (cf. \ref{sec:sm1} for details).

\subsection{Simulations}

Single-node basin stabilities and survivabilities have been estimated for all nodes in an ensemble of $M=50$ randomly generated networks with $N=100$ nodes each. Within each network half of the nodes act as net generators ($P=+1$), while the remaining nodes are net consumers ($P=-1$) such that there is an overall power balance, $\sum_{i=1}^{N}P_i=0$. Even though the synthetic power grid topologies generated by the random growth model are spatially embedded, the coupling strengths of the transmissions lines have been chosen uniformly to $K=6$ for simplicity and to highlight purely topological effects. The damping coefficients have been set uniformly to $\alpha=0.1$. Note that for this choice of parameters the preconditions for the approximation of the limit cycle position, equation (\ref{eqn:omega-LC}), are met.

\section{Results}
\label{sec:results}

\subsection{Interrelations between a node's stability and its topological properties}
\label{sec:stabMeas-vs-netMeas}

Firstly, we investigated interrelations between the node-wise stability measures and topological properties of the perturbed node. We highlight topological effects in the example of a node's degree $d$ and shortest-path betweenness $b$ \cite{Boccaletti2006}. They turned out to reveal the most prominent insights for transient and asymptotic stability, respectively, and are basic established local/mesoscale network characteristics.
Furthermore, this choice facilitates a comparison with previous findings \cite{Menck2014,Schultz2014}.

\subsubsection{Basin Stability}

We  regard different maximal perturbation levels $\Delta\omega=2.5$ to $12.5$, and observe that as expected the mean basin stability of all $N\cdot M=5000$ nodes decreases with growing perturbation levels $\Delta\omega$.

\begin{figure}[!ht]
	\centering
	\includegraphics[width=0.495\textwidth]{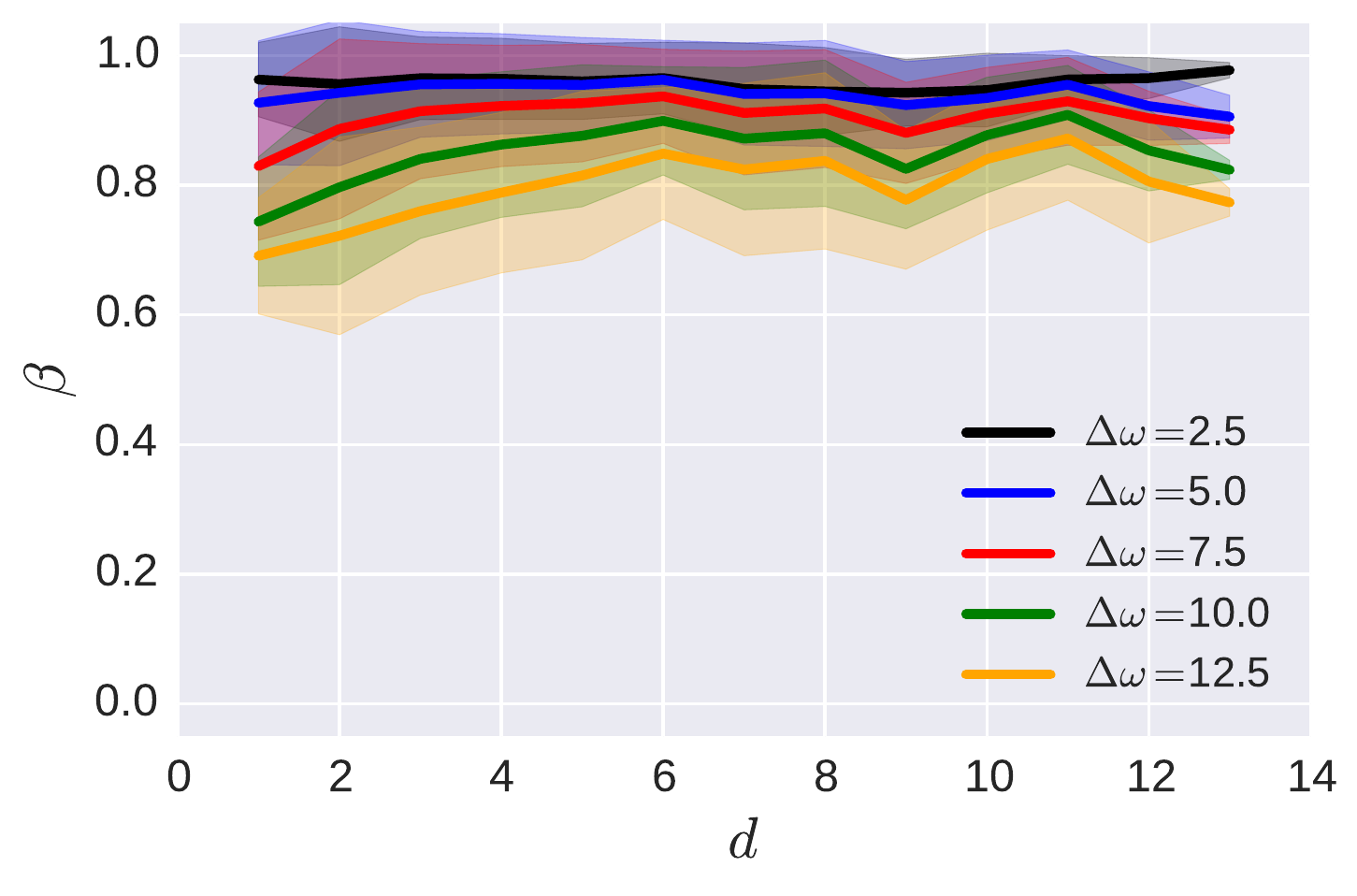}
	\includegraphics[width=0.495\textwidth]{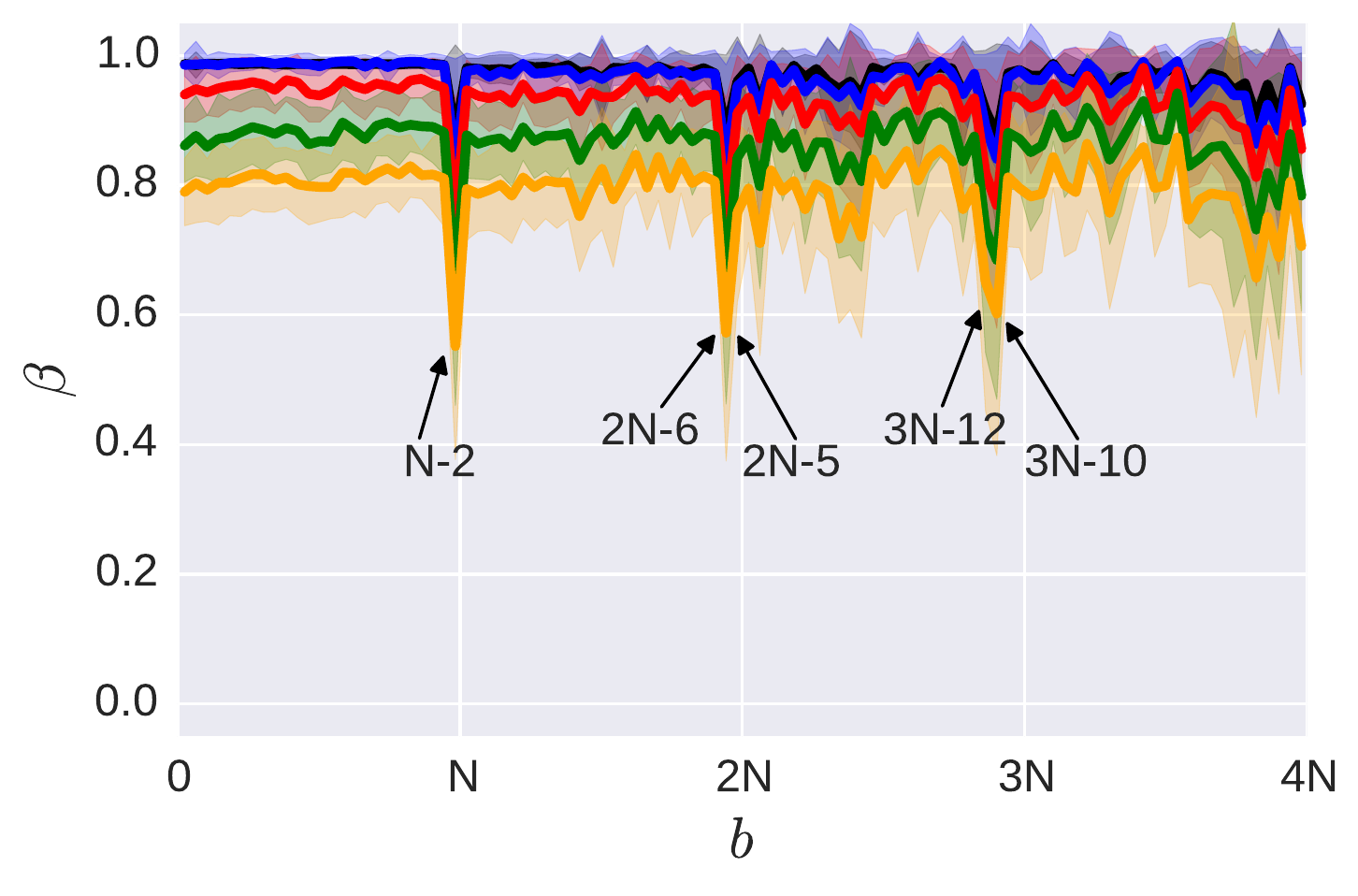}
	\caption[]{Dependence of single-node basin stabilities $\beta$ on the node's degree  $d$ (left) and betweenness  $b$ (right) for different levels of perturbations $\Delta\omega$. While there is no significant dependence for the degree, the basin stability features distinctive down-peaks at characteristic betweenness values which correspond to nodes situated within tree-shaped parts of the network. Bold lines show the means for a suitable binning of the data, shades indicate one standard deviation. Note that overlapping shades' colours can change.}
	\label{fig:BS-vs-network-measures}
\end{figure}

In order to detect which topological node characteristics influence asymptotic stability, the basin stability scores are regarded in dependence of the degree $d$ of the node which is defined as the number of neighbouring nodes (Figure \ref{fig:BS-vs-network-measures} a). While there is no significant dependency for lower perturbations levels ($\Delta\omega\leq5.0$) there is a slight increase of $\beta$ with $d$ for larger perturbation levels, for which there are, however, relatively large standard deviations in the stability estimates. Hence, degree alone is a weak predictor for basin stability which is in line with previous findings \cite{Menck2014, Schultz2014}.

There is, however, a characteristic dependence of basin stability on the betweenness $b$ of the node, which is defined as the number of shortest paths between all pairs of nodes within the network which are passing through the regarded node (Figure \ref{fig:BS-vs-network-measures} b). While there is no general trend for any perturbation level, distinctive down-peaks of basin stability are observable at certain betweenness values (as illustrated by Menck et al. \cite{Menck2014}). These particular values of $b$ correspond to nodes which lie within tree-shaped parts of the network (see the green-coloured nodes of the ``inner tree node'' category shown in Figure \ref{fig:example-network}). A similar dependency was found in \cite{Menck2014} for a maximal perturbation level of $\Delta\omega=100$. Hence we were able to qualitatively reproduce these findings for considerably smaller perturbation levels which appear more realistic by comparison to real-world cases.

\subsubsection{Survivability}

\begin{figure}[!ht]
	\centering
	\includegraphics[width=0.495\textwidth]{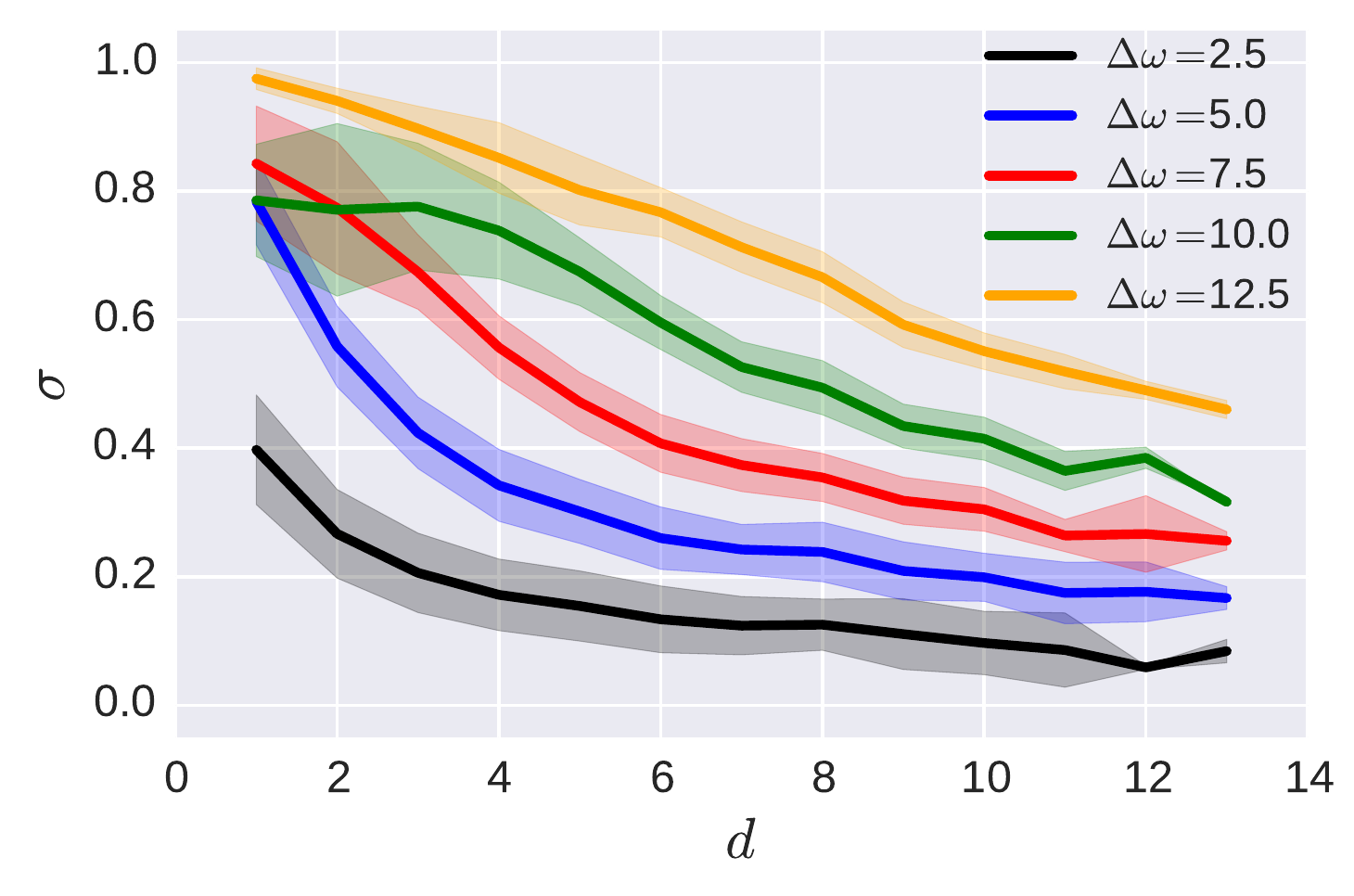}
	\includegraphics[width=0.495\textwidth]{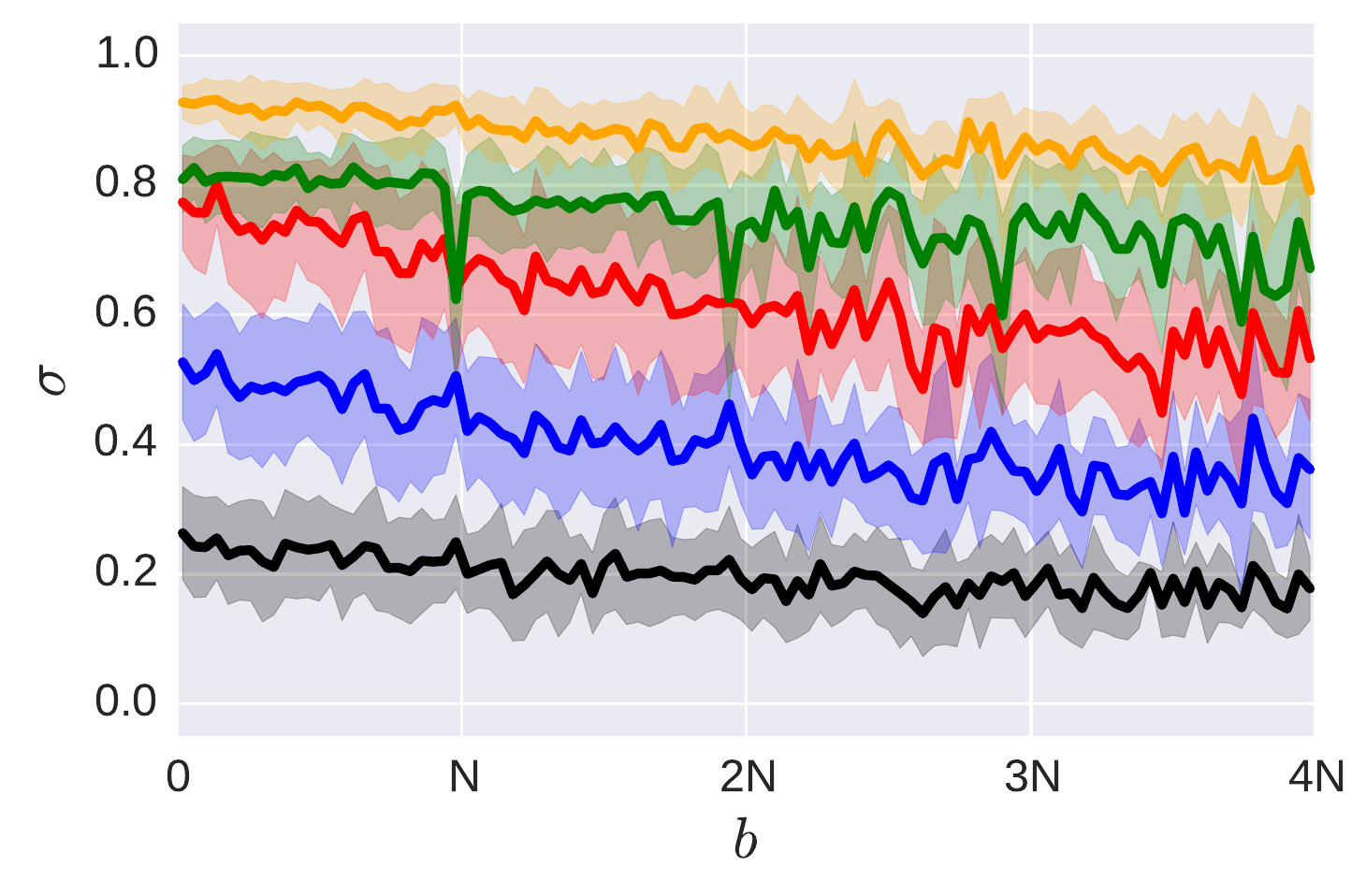}
	\caption[]{Dependence of single-node survivabilities $\sigma$ on the node's degree  $d$ (left) and betweenness  $b$ (right) for different levels of perturbations $\Delta\omega$. Both Figures show a negative correlation between a node's transient stability as measured by survivability and its topological properties within the network. This trend is most significant for the degree measure which makes it a suitable predictor of survivability, independent of the perturbation strength. Bold lines show the means for a suitable binning of the data, shades indicate one standard deviation. Note that the fluctuations of the mean lie within the one-standard-deviation-band, except for some characteristic values of $b$ indicated in Figure \ref{fig:BS-vs-network-measures}.}
	\label{fig:SV-vs-network-measures}
\end{figure}

Next, we studied the transient stability against large perturbations as measured by single-node survivability, again for different values of the maximal perturbation strength $\Delta\omega$ (Figure \ref{fig:SV-vs-network-measures}). The generally larger survivability scores achieved with increasing perturbations can be explained by the fact that the desirable region boundaries $\omega^+$ are increased simultaneously with $\Delta\omega$ (cf. Section \ref{sec:methods}). Another convention, which is not followed in this study, would be to hold the desirable region ($\omega^+$) constant when increasing the perturbations ($\Delta\omega$). In this case the average survivability of all nodes would decrease since for $\Delta\omega>\omega^+$ a certain fraction of trajectories would start outside $X^+$ and hence could not ``survive''.

In contrast to basin stability, the single-node survivability features a significant negative correlation with the degree which is found for all levels of perturbations (Figure \ref{fig:SV-vs-network-measures} a). This means that the frequency deviations within the network following a perturbation tend to be larger when nodes with a high degree are hit, thereby making the dynamics leave the desirable region. For example, for $\Delta\omega=12.5$ the probability of the trajectory to survive a perturbation is close to 1 if a node with degree $d=1$ is hit, while it lies below $0.5$ for nodes with degree $d=13$. For smaller perturbation levels the same trend is observable. Hence the degree may serve as a suitable predictor for the node's survivability.

For the betweenness measure the same but less striking trend is found (Figure \ref{fig:SV-vs-network-measures} b). Survivability is generally decreasing with the node's betweenness. Inner tree nodes which are characterized by specific values of $b$ do not stick out as dominantly as it was the case for basin stability. For lower perturbations ($\Delta\omega\leq5.0$) we observe small up-peaks in $\sigma$ at the specific $b$-values while for $\Delta\omega=10.0$ the same down-peaks as with basin stability occur. This reveals that nodes adjacent to tree-like structures are also crucial for predicting survivability, however, in a more subtle way compared to basin stability. Overall, betweenness alone is only a weak predictor for survivability, showing different features at different perturbation levels.

As for basin stability comparing survivability to other nodal network measures did not lead to more insights. Instead, a direct comparison of basin stability and survivability of the nodes was found to reveal non-trivial aspects of the system's dynamics and helped in identifying the node classification scheme introduced in \ref{sec:node-classification}.

\subsection{Joint distributions of basin stability and survivability}
\label{sec:stabMeas-joint}

\begin{figure}[!ht]
	\centering
	\includegraphics[width=0.495\textwidth]{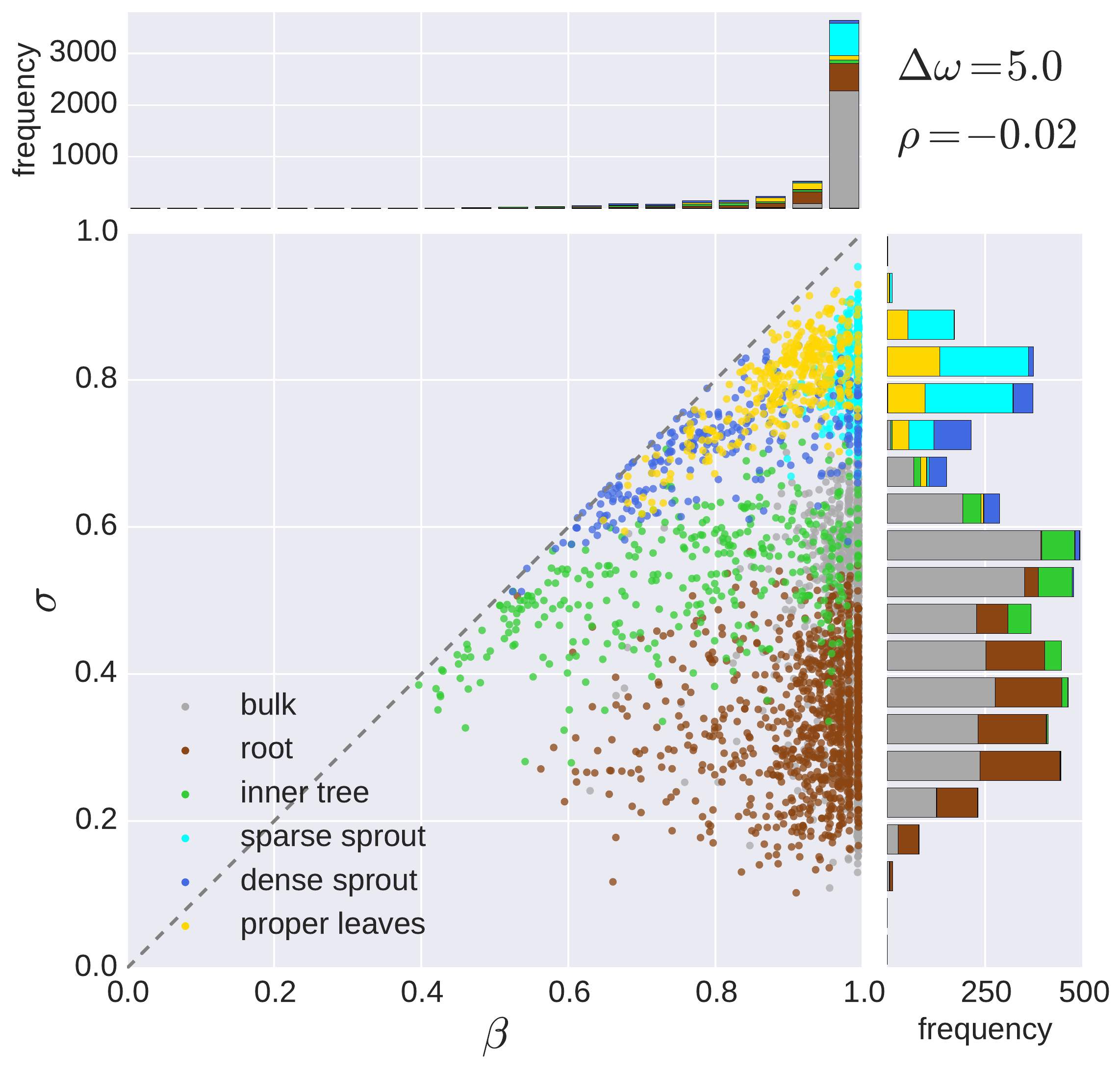}
	\includegraphics[width=0.495\textwidth]{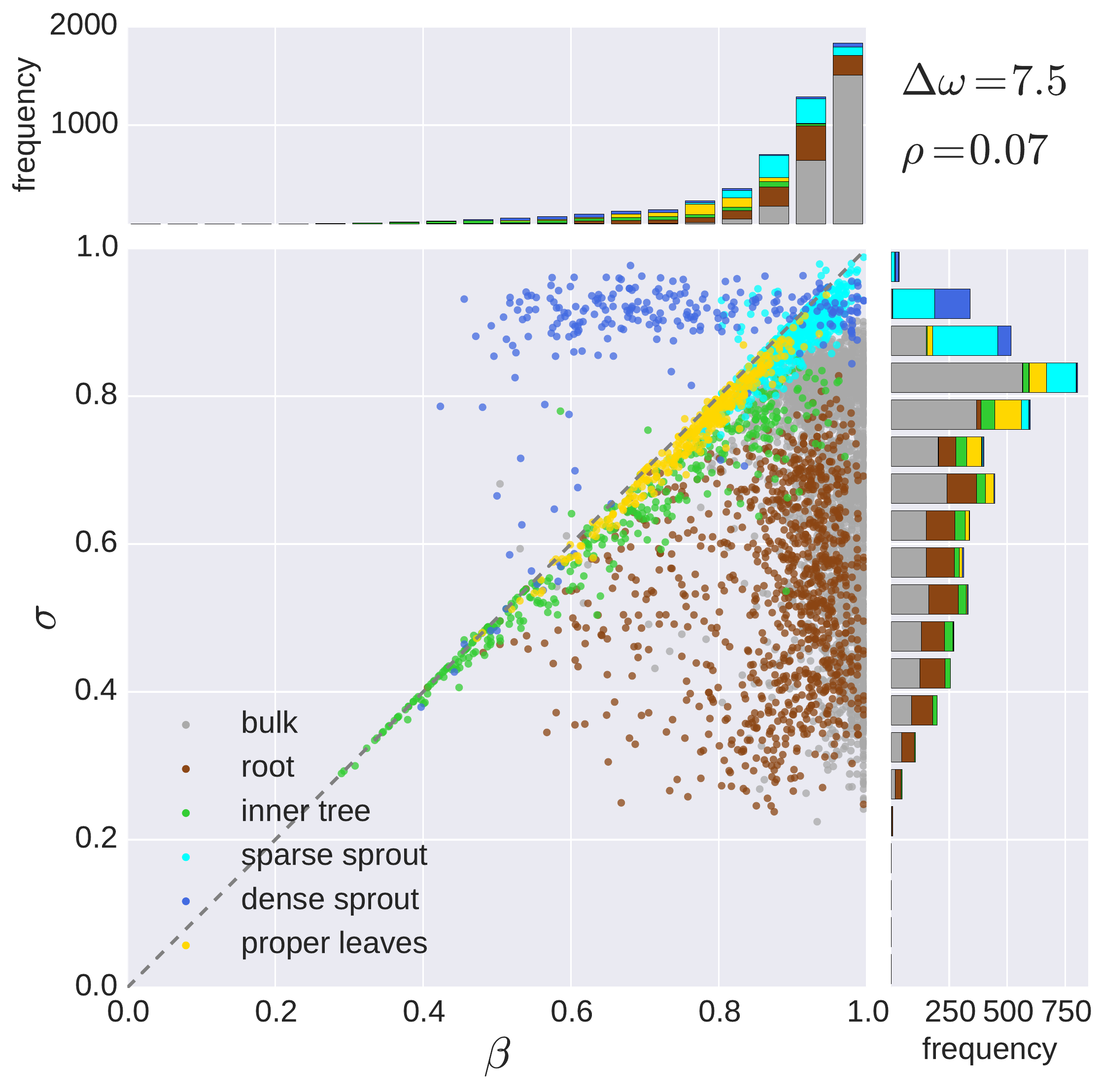}\\
	\includegraphics[width=0.495\textwidth]{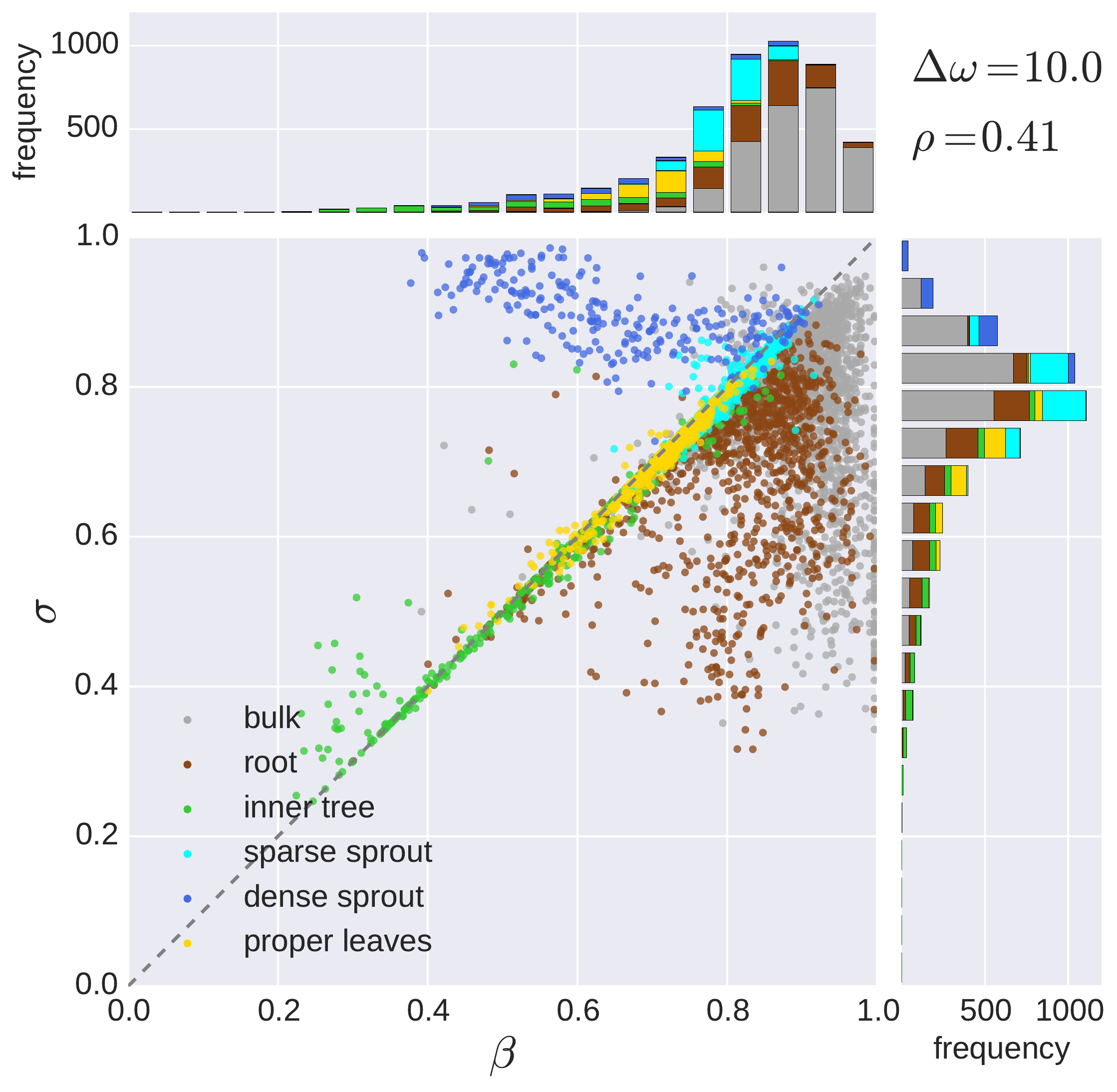}
	\includegraphics[width=0.495\textwidth]{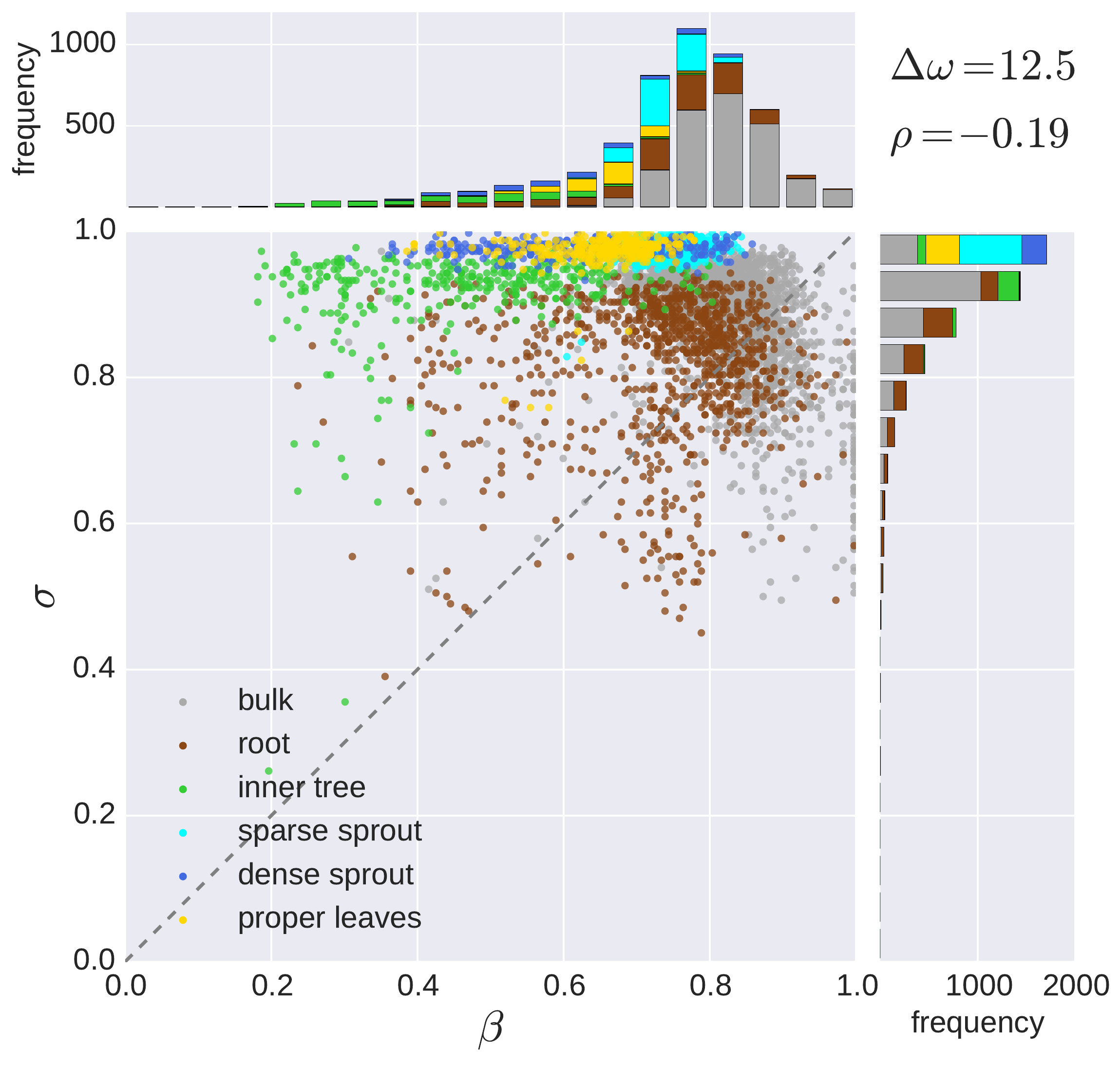}
	\caption[]{Scatter plots and distributions of single-node basin stabilities and survivabilities of all $N\times M=5000$ nodes for different perturbations levels $\Delta\omega$. The data points and bins are coloured according to the topological classification scheme introduced in Section \ref{sec:node-classification} and illustrated in Figure \ref{fig:example-network}. $\rho$ denotes the Pearson product-moment correlation coefficient.}
	\label{fig:SV-vs-BS-grouped}
\end{figure}

While this section focuses on the general distribution patterns of basin stability and survivability, the stability characteristics of the different node classes is described in the next section (\ref{sec:stabMeas-classes}). In order to reveal more insights about the interdependencies of asymptotic and transient dynamics as well as the relation to the network topology, we plotted the joint distribution of single-node survivability $\sigma$ and single-node basin stability $\beta$ for different perturbation levels $\Delta\omega$ (Figure \ref{fig:SV-vs-BS-grouped}). Each panel shows a total of $N\times M=5000$ individual estimates.

Let us focus on the two marginal distributions first. For a rather low maximal perturbation level ($\Delta\omega=5.0$) the distribution of basin stability is extremely skewed with $73.0\,\%$ of nodes featuring $\beta\geq0.95$. The survivabilities in turn show a widely spread bimodal distribution. This fact shows that survivability is generally much more influenced by topology than basin stability at lower perturbation levels.

At larger perturbation levels ($\Delta\omega\geq7.5$) the distributions of both basin stability and survivability are unimodal and skewed. For $\Delta\omega=7.5$ still $62.4\,\%$ of nodes feature a high asymptotic stability of $\beta\geq0.9$. The distribution of transient stability is still wide but shifted towards larger values. While some groups of nodes show a strong correlation between $\beta$ and $\sigma$, the overall Pearson correlation coefficient is close to zero ($\rho=0.07$). 

This picture is reversed when looking at the highest perturbation level of $\Delta\omega=12.5$. Here $62.9\,\%$ of nodes have survivabilities of $\sigma\geq0.9$ and the distribution of basin stability is rather widespread. It is known that at considerably larger perturbation levels ($\Delta\omega=100$) also basin stabilities show a widespread multimodal distribution \cite{Menck2014, Schultz2014a}. Hence basin stability might be a more useful measure for the system's stability if very large perturbations are expected.

The intermediate case with $\Delta\omega=10.0$ is particularly interesting. Here the distributions of $\beta$ and $\sigma$ are very similarly shaped and numerous nodes show a strong correlation. However, there are also many nodes for which there is no such correlation, with the overall Pearson correlation coming out at $\rho=0.41$.

\subsection{Basin stability and survivability for different classes of nodes}
\label{sec:stabMeas-classes}

The findings presented so far are in line with the reasoning of Hellmann et al. who state that basin stability and survivability are generally not correlated and hence the asymptotic behaviour does not allow conclusions about the transient one \cite{Hellmann2016}.
We now want to achieve a more precise statement by putting a stronger focus on the patterns observed in the joint distribution of the stability measures in Figure \ref{fig:SV-vs-BS-grouped}.

In order to decipher how these patterns relate to topological characteristics, it turned out to be helpful to focus on nodes located inside or adjacent to tree-shaped parts of the network and to distinguish several types of nodes by how far inside the tree they are located (cf. Section \ref{sec:node-classification} and Table \ref{tab:overview-classification}). This is also suggested by the findings presented in Section \ref{sec:stabMeas-vs-netMeas}.

Each of the previously defined classes features typical characteristics regarding asymptotic and transient stability at different perturbation levels (Figure \ref{fig:SV-vs-BS-grouped}). We begin with discussing $\Delta\omega=5.0$. In this case, for all nodes the survivability is lower than the basin stability. This indicates that there are no undesirable attractors within the survival region $X^+$. Hence basin stability sets an upper limit to survivability. All trajectories which stay within the desirable region $X^+$ have to converge to the synchronous operating state of the grid, $X^\star$ \cite{Hellmann2016}. This is expected as a desynchronization is expected to lead to oscillators rotating with their natural frequency, which here is $\omega_\text{LC}=P/\alpha=10.0$ (cf. (\ref{eqn:omega-LC})).

At rather low perturbation levels ($\Delta\omega=5.0$) the nodes forming the upper mode of the survivability distribution are the leaves. Dense sprouts show lower survivabilities than sparse sprouts, indicating that also a low neighbour degree is beneficial for survivability. The lower mode of the distribution is formed by bulk, root and inner tree nodes. As expected from the degree dependence (cf. Figure \ref{fig:SV-vs-network-measures}) inner nodes show higher survivabilities than root nodes which are most critical. Hence, the transient dynamics following a perturbation of a root node tend to exhibit large frequency deviations, leading to a transgression of the boundaries of the desirable region.

For $\Delta\omega\geq7.5$ we observe surprising new behaviour. While perturbations at most nodes still behave as if there was no undesirable attractor in the survival region, perturbations originating at a dense sprout almost all have $\sigma>\beta$. This means a novel asymptotic state or a very long transient, is reached, in which the system is not synchronized, but as the system does not leave the survival region $X^+$, no node is swinging at its natural frequency $\omega_\text{LC}$ either. Exemplary trajectories for perturbations at a dense sprout are shown in Figure \ref{fig:trajectories} in \ref{sec:supp-figs} and reveal that we indeed observe a novel asymptotic state, with a solitary desynchronized oscillator not swinging at its natural frequency.
To our knowledge, this state has not previously been observed in these systems, for example in the bifurcation studies of \cite{Olmi14, Olmi15}. Indeed we expect that this state would be very hard to observe, if the initial conditions are drawn fully randomly, and not localized at individual nodes.

For sparse sprouts, proper leaves and inner tree nodes $\beta$ and $\sigma$ are strongly positively correlated. This means that for these nodes trajectories which leave the desired region tend to converge to non-synchronous states. In other words, the basin of attraction of the synchronized state is entirely contained in $X^+$. Root nodes show a pattern contrary to dense sprouts with a wide range of survivabilities and rather high basin stabilities.

The patterns yielded at $\Delta\omega=10.0$ are very similar to those at $\Delta\omega=7.5$. As the frequency of the limit cycle fluctuates around $10.0$, drawing the boundary exactly there means that still all nodes that fully desynchronize, and go to their natural frequency, will hit the survivability threshold.  Now a few nodes besides the dense sprouts feature $\sigma>\beta$. A further anomaly that can be observed here is that the mean survivabilities of the leaves are smaller for $\Delta\omega=10.0$ than for $\Delta\omega=7.5$, opposed to the general observation of increasing survivabilities for larger values of $\Delta\omega$.

Finally, at $\Delta\omega=12.5$, the natural frequency of the oscillators is fully within the survival region, and a desynchronized node is characterized as having survived. Consequently, the majority of nodes has larger survivabilities than basin stabilities. There are numerous non-synchronous states whose trajectories lie completely within the desirable region $X^+$. The limit cycle trajectory is hence also prominent in the multi-node network model. While the leaves of the tree-like parts show a similar pattern in the $\sigma$-$\beta$-space, the inner tree nodes are clearly separated at slightly lower survivabilities and fairly lower basin stabilities. Nodes from the bulk feature the largest basin stabilities.

Independent of the perturbation level $\Delta\omega$, dense sprouts tend to feature lower basin stabilities than sparse sprouts. Menck et al. studied the dependence of basin stability on the (average) degree of the neighbour(s) for all leaves (``dead ends'') but did not find a significant correlation \cite{Menck2014}. It is our finer partition of leaves into proper leaves and sprouts, the neccesity of which was recognized by studying the joint plots, that allows more detailed insights.

Complementary to
Figure \ref{fig:SV-vs-BS-grouped}, the partition achieved using the defined topological classes of nodes is highlighted in three-dimensional representations of the joint distributions of basin stability and survivability at different perturbation levels (Figure \ref{fig:scatter3D} in \ref{sec:supp-figs}).

\section{Discussion}
\label{sec:discussion}

\subsection{Relevance for designing stable power grids}
The insights gained from this study are particularly relevant for both the stability assessment and the design of stable power grids. The employed nonlinear stability measures, basin stability and survivability, provide useful information on the asymptotic and transient stability of a grid against large perturbations, thereby surpassing the insights gained from local, linear stability assessments. The probabilistic measures have the benefits of being intuitive to understand and efficient to estimate for the overall system, irrespective of the complexity of the dynamics. They thus allow, for the first time, a systemic understanding of dynamical effects in the overall system.

Of particular relevance to the systemic stability of the power system are the novel asymptotic states we discovered, which primarily arise from perturbations at dense sprouts. The perturbed nodes are desynchronized and oscillate at a frequency much smaller than their natural. They are thus a pure network effect that can not be understood by studying individual machines. If such desynchronized states exist within sufficiently narrow frequency bounds, they would be a severe systemic risk to the power grid.
They might be related to observed phenomena like Inter Area Oscillations, in which a deviation from perfect synchrony is observed. 
While an extensive analysis of these states is not part of this work, the topological characterization of dynamic systems already suggests a mean to prevent them preemptively in the design of the power grid, by avoiding dense sprouts.

Beyond this novel dynamical phenomenon, our study revealed various interrelations between the pure topological properties of a node and its stability. Tree-like structures, which contain about half of the nodes in the network ensemble, were found to be characterized by stability properties which significantly distinguish them from the remaining bulk of nodes. We were able to identify a small set of topological groups whose nodes feature similar patterns of asymptotic and transient stability at various perturbation levels. The knowledge about these patterns can in turn be utilized to predict a node's relevance for the whole grid's stability knowing only its topological embedding within the network. Given a concrete real-world power grid topology, our classification scheme hence enables the identification of potentially vulnerable nodes. While outside the scope of this work we expect that it will be possible to extend this identification to more nodes by means of statistic regressions taking network measures as input \cite{Schultz2014a}.


Menck et al.\ found that the weak basin stability of inner nodes can be cured by reconnecting leaf nodes (there termed ``dead ends'') to the grid  \cite{Menck2014}. This agrees well with our observation above to avoid sprouts and that ``detour'' nodes are favoured by our stability measures \cite{Schultz2014a}. Another strategy suggested by our findings would be to avoid high-degree nodes as these feature the worst survivabilities, independent of the perturbation level. By combining the above-mentioned rules, reconnection of leaves and avoidance of hubs with high centralities, a possible design principle for stable power grids would be to strive for rather homogeneous network topologies, characterized by narrow degree distributions. Using linear stability techniques such homogeneous topologies have also be found to be generally easier to synchronize \cite{Nishikawa2003}, while tree-shaped structures apparently show rather bad synchronizability \cite{Yook2006}. Interestingly, while tree-shaped structures feature both poor linear (synchronizability) and nonlinear (basin stability, survivability) stability properties, in small-world topologies synchronizability and basin stability were found to behave contrarily \cite{Menck2013}.

It should be pointed out that for our analysis we employed conceptual models of the transmission grid. A direct transfer of the findings to lower grid levels might not be valid. Another simplifying assumption which does not hold for real-world grids is the uniform distribution of both the power injections and loads. Whether rather heterogeneous power distributions would affect the findings has not been investigated. We are also aware that the maximal perturbation levels assumed for our simulations are rather high. Instead of assuming uniformly distributed perturbations it would be more realistic to assume a unimodal distribution in which small perturbations in both the phase and frequency deviations are more frequent and large perturbations are rare events. Realistic distributions of perturbations could, for instance, be derived from the research on intermittency in power fluctuations from wind or solar power systems \cite{Anvari2016,Milan2013}. This work should thus be seen as an exploration of phase space properties like the structure of a state's basin of attraction, rather than a concrete study of realistic power grid models.



\subsection{Relevance for complex dynamical networks in general}

Beyond the direct implications for the resilience of power grids, our findings show the power of \textit{sampling-based} dynamical methods, like \textit{survivability} and \textit{basin stability}. These are evaluated by sampling the system's dynamical reaction to non-small perturbations, here in phase space, but more generally also in parameter space. Combined with network measures and topological classifications, they provide general insights for complex dynamical networks. We expect that the crucial role of tree-shaped parts for the dynamics found here, is not specific to power grids but might rather be a general phenomenon in oscillator networks, infrastructure networks or even biological networks, e.g., in neural or genomic dynamics. The suggested topological classification scheme and the terminology for nodes in trees are independent of the nodal dynamics and are thus easily applicable to networks of different types. 

As the employed synchronous machine model (or ``Swing Equation'') is equivalent to the Kuramoto model with inertia \cite{Rodrigues2016}, the findings are particularly relevant for the general study of oscillator networks. The combined analysis of transient and asymptotic behaviour via basin stability and survivability allows indirect insights into the geometry of the system's phase space. The identification of nodes for which $\sigma>\beta$ allowed us to infer the existence and position of new types of limit cycles. Particularly, for the group of dense sprouts we found that non-synchronous states exist besides those given by the approximated solution to the single-node system (\ref{eqn:omega-LC}).

In a first order approximation the coupling strength $K$ in (\ref{eqn:omega-LC}) is proportional to a node's degree $d$, while $P$ and $\alpha$ are independent of topological characteristics. Hence the approximation (\ref{eqn:omega-LC}) suggests that the \textit{amplitudes} $A_\text{LC}=PK/\alpha$ of a limit cycle are proportional to the degree of the respective non-synchronously rotating nodes. This is a contributing mechanism to the low survivability of high degree nodes for very large perturbations. We suspect that it will be possible to gain further understanding of the relationship between topology and dynamics through more sophisticated approximations. Understanding the dependencies of the asymptotic spectrum of networked dynamical systems on the ambient topology in more detail will however require significant further work, both numerical and analytical.

\section{Conclusions}
\label{sec:conclusion}

Our results form another step towards a better understanding of the interrelations between topology and stability in complex dynamical networks. Tree-shaped topologies which are particularly prominent in infrastructure networks, have been found to feature stability properties which considerably deviate from those of the remaining bulk of nodes. A topological classification scheme for nodes adjacent to those tree-shaped parts has been suggested which enables a prediction of a node's transient and asymptotic stability against large perturbations. This classification of nodes can hence aid both the stability assessment and the design of stable infrastructure systems.

The sampling based stability measures we employed were shown to enable surprising novel insights into the asymptotic dynamics of networked dynamical systems, revealing both, previously unknown asymptotic states and surprisingly precise relationships between the topology and these novel states.

Due to the parametrization of the model equations, the results are particularly relevant in the context of power grid research. If both high asymptotic stability (reflected by single-node basin stability) and transient stability (reflected by survivability) of power grids are desired, avoiding both sparsely connected tree-shaped structures and high-degree hub nodes appears to be a promising design principle.

Independent of the particular application the presented study shows how the nonlinear stability concepts of basin stability and survivability can be combined to gain a better understanding of a -- not necessarily networked -- dynamical system.





\ack

The authors gratefully acknowledge the support of BMBF, CoNDyNet, FK. 03SF0472A.

The authors gratefully acknowledge the European Regional Development Fund (ERDF), the German Federal Ministry of Education and Research and the Land Brandenburg for supporting this project by providing resources on the high performance computer system at the Potsdam Institute for Climate Impact Research.


We further thank Peng Ji for helpful discussions regarding the interpretation of the results.

\section*{References}
\bibliographystyle{my-iopart-num}
\bibliography{references}


\appendix

\section{Algorithm for the identification and classification of nodes in tree-shaped parts of networks}
\label{sec:sm2}

Let $G=(V,E)$ be an undirected graph that is connected and not itself a tree (i.e., contains at least one cycle).
Our goal is to identify all nodes $x\in V$ that are in a tree-shaped part of $G$ 
and classify them using height and depth.
As defined in the main text, a \textit{tree-shaped part} of $G$ 
is an induced subgraph $T'=(V',E')=G|_{V'}$ of $G$ 
(i.e., a subset of nodes $V'\subseteq V$ together with the set $E'$ of \textit{all} edges in $E$ between nodes in $V'$) 
that is maximal with the property that there is exactly one node $r\in V'$ that has at least one neighbour in $G-T'$.
$r$ is then called the \textit{root} of $T'$, and one can see easily that it must have degree at least three.
For any graph $G'=(V',E')$ and node $x\in V'$, we denote by $d_{G'}(x)$ the degree of $x$ in $G'$. 
A node with $d_{G'}(x)=1$ is called a \textit{leaf} of $G'$.

A simple algorithm to identify all tree-shaped parts of $G$, their roots, 
and the parents, children, heigths, depths, and branches of all their members
is the following. 

In the first part, we iteratively define
\begin{itemize} 
  \item a decreasing sequence of node sets $V_0\supset V_1\supset V_2\ldots$,
  \item the respective induced subgraphs $G_i = G|_{V_i}$,
  \item a sequence of disjoint height level sets $H_i$,
  \item parents $\pi(x)$,
  \item sets of children $C(x)$,
  \item branches $B(x)$,
  \item and height labellings $\eta(x)$,
\end{itemize}
by successively removing leaves from the remaining graph as follows.
Put $V_0 := V$ and initially $C(x) := \emptyset$ for all $x\in V$. 
Given $V_i$ and $G_i := G|_{V_i}$,
let $H_i := \{x\in V_i: d_{G_i}(x) = 1\}$ be the set of leaves of $G_i$.
For each $x\in H_i$, 
let the parent of $x$, $\pi(x)$, be the unique neighbour of $x$ in $G_i$;
add $x$ to its set of children, $C(\pi(x))$.
Note that $\pi(x)\in V_i - H_i$.
The branch of $x$ is $B(x) := \{x\} \cup \bigcup_{y\in C(x)} B(y)$,
and the height is $\eta(x) = i$.
As long as $H_i \neq \emptyset$, put $V_{i+1} := V_i - H_i$ and repeat.

To finishing the first part after these iterations, 
let $N := \bigcup_i H_i$ be the set of all thus identified non-root nodes,
let $R := \{\pi(x): x\in N\} - N$, and call each $r\in R$ a \textit{root}.
Put $B(r) := \{r\} \cup \bigcup_{y\in C(r)} B(y)$ and
$\eta(r) := 1 + \max\{\eta(x): x\in N,\pi(x)=r\}$ for all $r\in R$.
The tree-shaped parts $T'$ of $G$ are now exactly the subgraphs $T'=G|_{B(r)}$ induced by the branches of any roots $r\in R$.

In the second part,
we define a \textit{depth} $\delta(x)$ for each $x\in N\cup R$, counted outwards starting from the roots, 
in addition to the height, which is counted inwards starting from the leaves.
This is again done iteratively by defining a sequence of disjoint depth level sets $D_i$.
Put $D_0 := W_0 := R$, and put $\eta(x) := 0$ for each $x\in D_0$.
Having defined $D_{i-1}$ and $W_{i-1}$, 
define $D_i := \bigcup_{x\in D_{i-1}} C(x) - W_{i-1}$
and $W_i := D_{i-1} \cup D_i$,
and put $\delta(x) := i$ for each $x\in D_i$,
iterating this until $D_i = \emptyset$.
Note that $\delta(x)$ is the distance from $x$ to the root of its tree-shaped part.

Finally, we put
$S :=\left\{ x\in N \,\vert\, \eta(x)=0 \wedge \delta(x)=1 \right\}=\left\{ x\in L \,\vert\, \delta(x)=1 \right\}$ (sprouts),
$S_d :=\left\{ x\in S \,\vert\, \bar{d}_\mathcal{N}>5 \right\}$ (dense sprouts), 
$S_s :=\left\{ x\in S \,\vert\, \bar{d}_\mathcal{N}<6 \right\}$ (sparse sprouts),
$P :=\left\{ x\in N \,\vert\, \eta(x)=0 \wedge \delta(x)>1 \right\}=\left\{ x\in L \,\vert\, \delta(x)>1 \right\}$ (proper leaves).

\section{Estimation of basin stability and survivability from simulations}
\label{sec:sm1}

Both employed stability measures, basin stability $\beta$ (\ref{eqn:muB}) and survivability $\sigma$ (\ref{eqn:muS}), have been estimated using Monte-Carlo sampling. For each of the $M\times N=5000$ nodes $L=200$ trajectories with perturbed initial conditions (\ref{eqn:phi-pert}) and (\ref{eqn:omega-pert}) have been simulated.

If $s$ of these trajectories return (sufficiently close) to $X^\star$ after the simulation time of $t=100$, their fraction is used as an estimator of $\beta$:
\begin{eqnarray}
	\hat{\beta}=\frac{s}{L}	\label{eqn:muB-est}
\end{eqnarray}
Since the perturbed trajectories either converge or not, the sampling of initial conditions can be regarded as a Bernoulli experiment. Thus the standard error of the probability estimator is given by
\begin{eqnarray}
	e_{\hat{\beta}}=\sqrt{\frac{\hat{\beta}(1-\hat{\beta})}{L}}	\label{eqn:mu-standard-error}
\end{eqnarray}
If either all or none of the perturbed trajectories converge the estimated standard error becomes zero which is troublesome for further statistical calculations. Hence we used the Agresti-Croull method for a better estimation of $\beta$ and its standard error $e_{\beta}$ \cite{Agresti1998}. For the desired confidence
this corresponds to adding one trial which is ``half success'' and ``half failure''. Defining $\tilde{s}=s+\frac{1}{2}$ and $\tilde{L}=L+1$, the corrected estimator $\tilde{\beta}$ is given by
\begin{eqnarray}
	\tilde{\beta}=\frac{\tilde{s}}{\tilde{L}}		\label{eqn:muB-est-AC}
\end{eqnarray}
and the corresponding standard error amounts to
\begin{eqnarray}
	e_{\tilde{\beta}}=\sqrt{\frac{\tilde{\beta}(1-\tilde{\beta})}{\tilde{L}}}
\end{eqnarray}
Analogously, $\sigma$ and its standard error $e_\sigma$ have been estimated using the fraction of trajectories which did not leave the desirable region $X^+$ given by (\ref{eqn:X+}) within the simulation time.

\section{Supplementary Figures}
\label{sec:supp-figs}

\begin{figure}[!ht]
	\centering
	\includegraphics[width=0.8\textwidth]{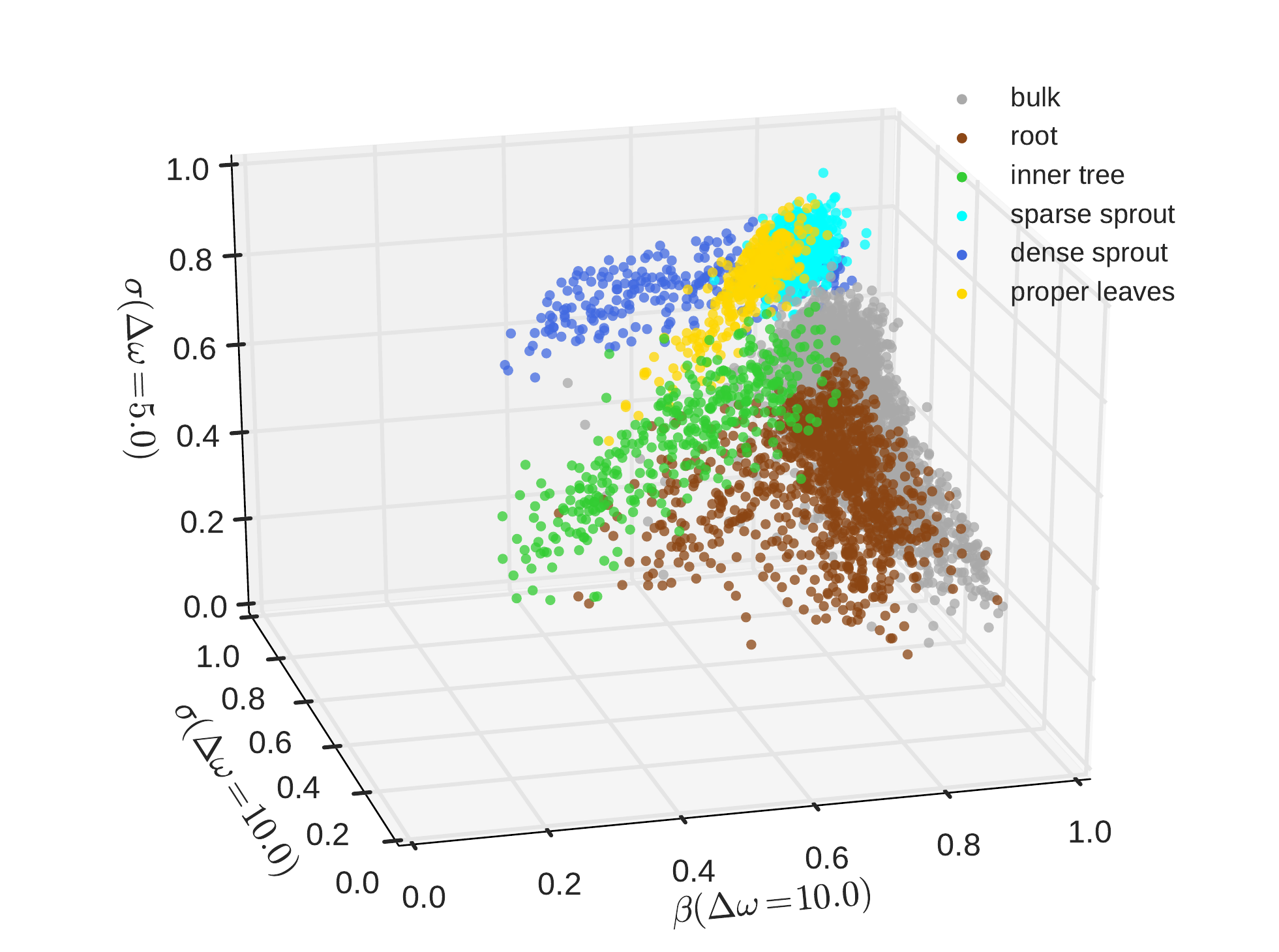}
	\includegraphics[width=0.8\textwidth]{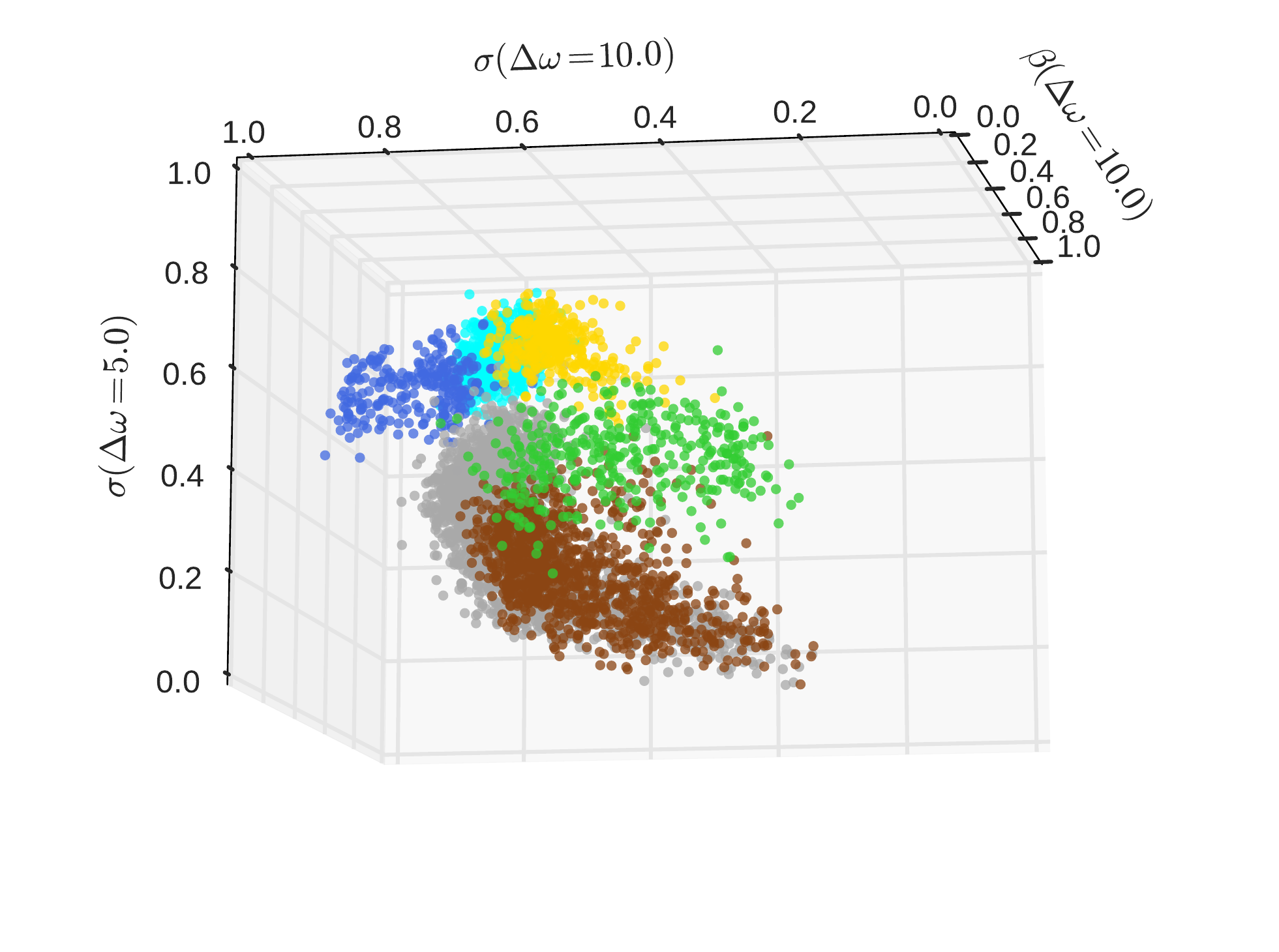}
	\caption[]{Three-dimensional scatter plots of single-node basin stabilities $\beta$ (for $\Delta\omega=10.0$) and survivabilities $\sigma$ (for $\Delta\omega=5.0$ and $\Delta\omega=10.0$) from two different view angles. The data points are coloured according to the topological classification scheme introduced in Section \ref{sec:node-classification} and illustrated in Figure \ref{fig:example-network}. The three-dimensional representation of the data shows the clear separation of the topological classes of nodes with respect to their asymptotic and transient stability properties.}
	\label{fig:scatter3D}
\end{figure}

\begin{figure}[!ht]
	\begin{center}
	\includegraphics[width=0.48\textwidth]{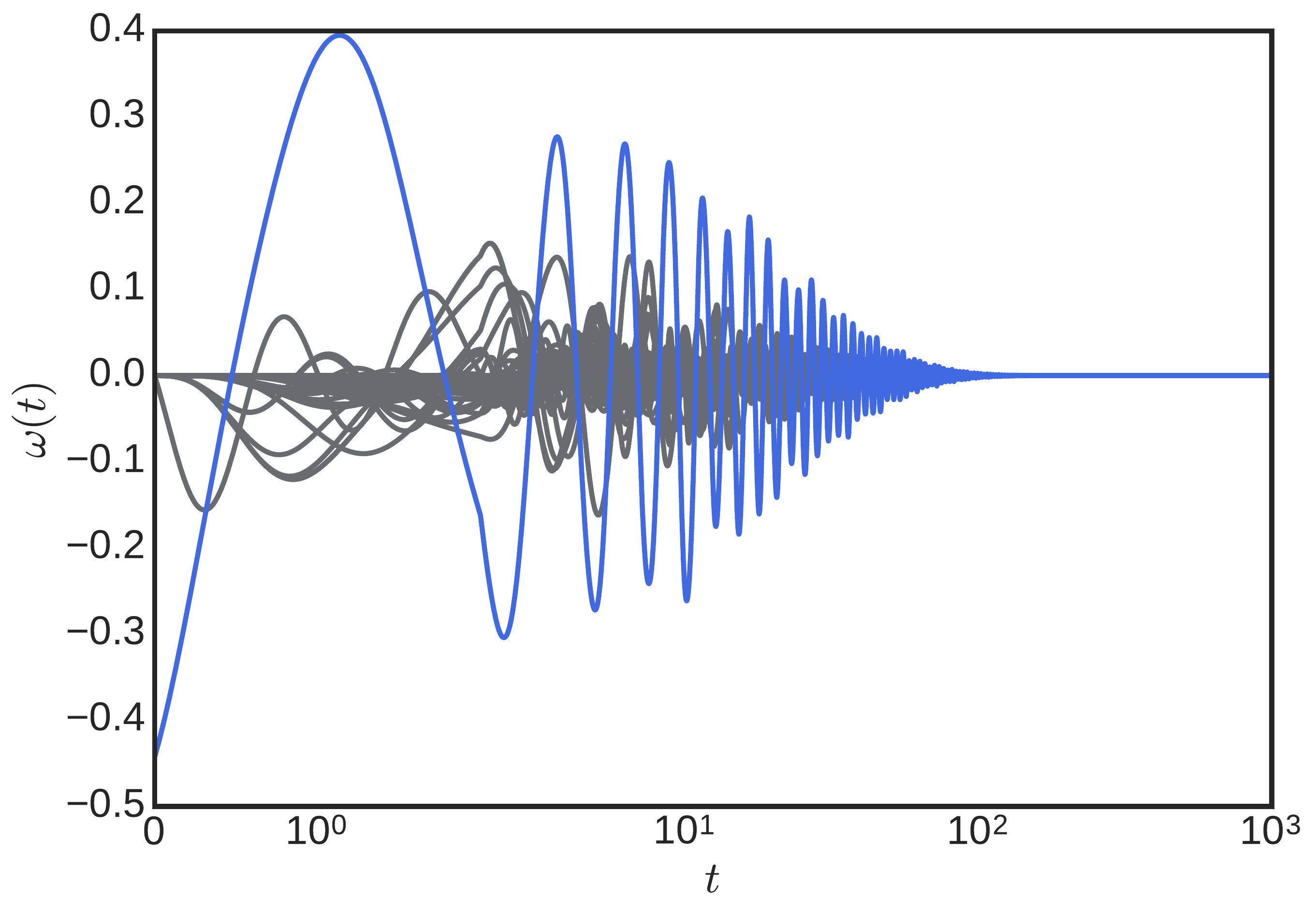}
	\end{center}
	\includegraphics[width=0.48\textwidth]{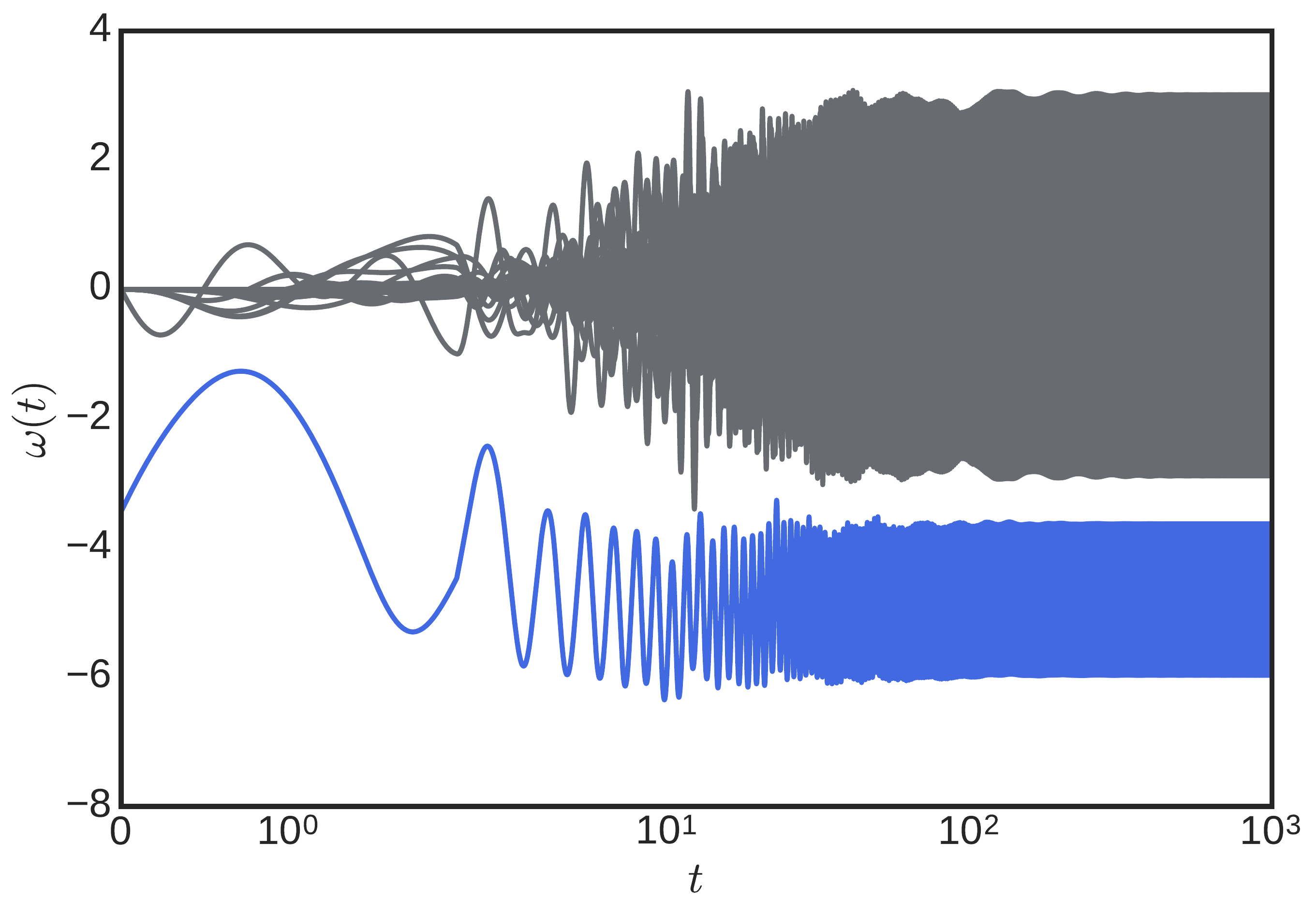}
	\includegraphics[width=0.48\textwidth]{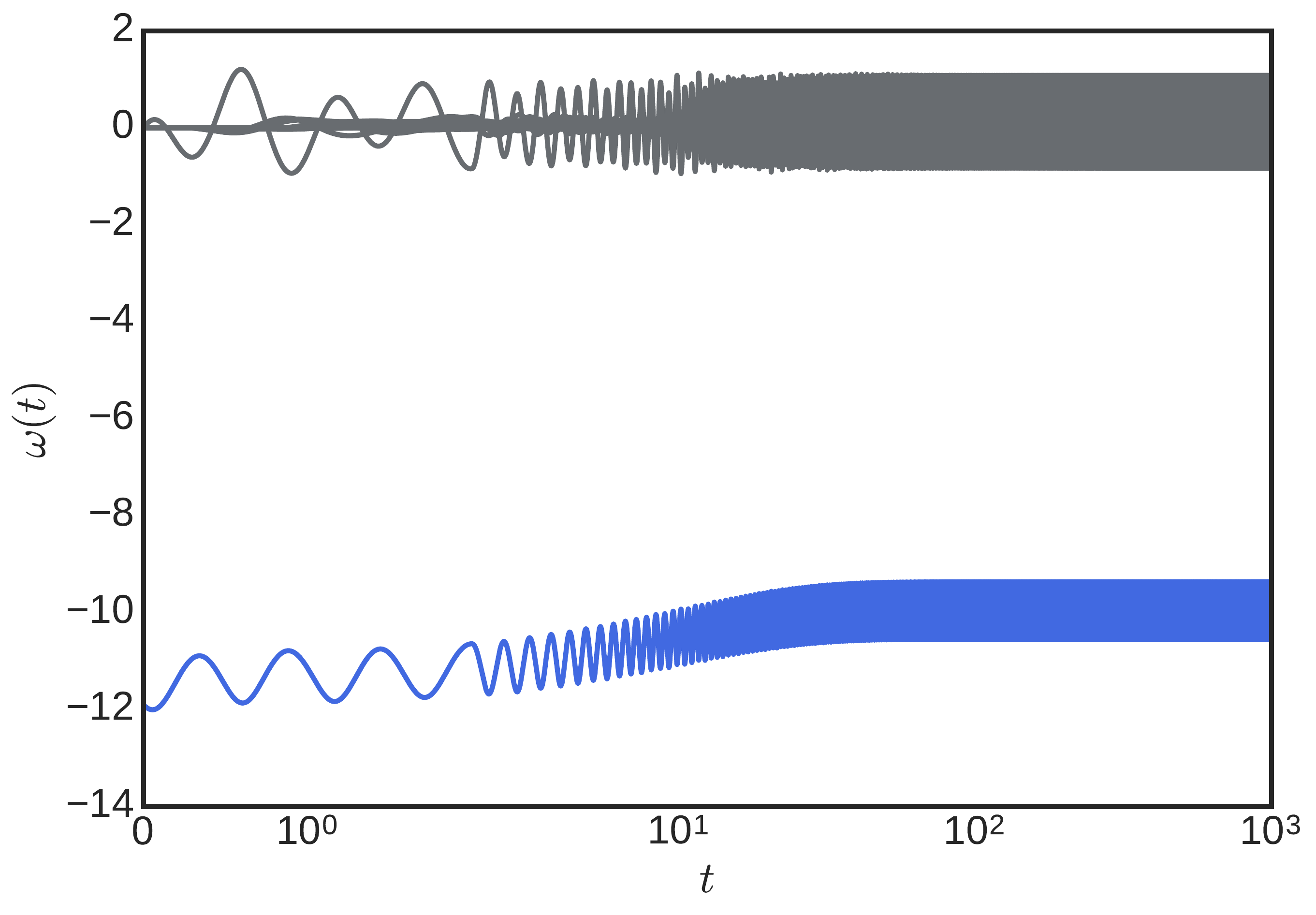}
	\caption[]{Exemplary trajectories of the frequency deviations $\omega(t)$ for perturbations at a dense sprout which are representative for the different observed asymptotic states. The trajectory of the perturbed node is coloured blue while those of all other nodes in the network are shown in grey.
	For low perturbations (upper panel) the frequency deviation of the perturbed node declines over a long transient phase until the network settles to the synchronous state after about 100 seconds.
	At medium perturbations levels (lower left panel) we observe the novel asymptotic state for which the perturbed node oscillates around a frequency deviation of about half of its natural frequency, while relatively strong oscillations around $\omega=0$ are observed in the remaining network.
	Finally, for rather high perturbations (lower right panel) the dense sprout asymptotically oscillates around its natural frequency $\omega_\text{LC}=P/\alpha=10.0$, as suggested by the approximation \ref{eqn:omega-LC}, while the remaining network shows oscillations around $\omega=0$ but with a smaller amplitude than in the previous case.
	Note that the impression of accelerating oscillations is due to the logarithmic time axis.}
	\label{fig:trajectories}
\end{figure}

\end{document}